\documentclass[preprint,times,tighten]{aastex631}

\usepackage{epsf,color,amsmath}

\usepackage{cancel}

\newcommand{\sfrac}[2]{\mathchoice%
  {\kern0em\raise.5ex\hbox{\the\scriptfont0 #1}\kern-.15em/
    \kern-.15em\lower.25ex\hbox{\the\scriptfont0 #2}}
  {\kern0em\raise.5ex\hbox{\the\scriptfont0 #1}\kern-.15em/
    \kern-.15em\lower.25ex\hbox{\the\scriptfont0 #2}}
  {\kern0em\raise.5ex\hbox{\the\scriptscriptfont0 #1}\kern-.2em/
    \kern-.15em\lower.25ex\hbox{\the\scriptscriptfont0 #2}} {#1\!/#2}}

\newcommand{\C}{$^{12}$C}

\newcommand{\castro}{{\sf Castro}}
\newcommand{\amrex}{{\sf AMReX}}
\newcommand{\pynucastro}{{\sf pynucastro}}
\newcommand{\microphysics}{{\sf Microphysics}}

\usepackage{comment}
\usepackage{ulem}

\begin{document}
\title{Sensitivity of He Flames in X-ray Bursts to Nuclear Physics}

\shorttitle{XRB Flame
Sensitivity}
\shortauthors{Chen et al.}

\author[0000-0002-2839-107X]{Zhi Chen}
\affiliation{Department of Physics and Astronomy,
Stony Brook University,
Stony Brook, NY 11794-3800, USA}

\author[0000-0001-8401-030X]{Michael Zingale}
\affiliation{Department of Physics and Astronomy, Stony Brook University,
             Stony Brook, NY 11794-3800, USA}

\author[0000-0001-6191-4285]{Kiran Eiden}
\affiliation{Department of Astronomy,
University of California, Berkeley,
Berkeley, CA 94720-3411, USA}



\correspondingauthor{Zhi Chen}
\email{zhi.chen.3@stonybrook.edu}

\begin{abstract}
Through the use of axisymmetric 2D hydrodynamic simulations, we further investigate laterally propagating flames in X-ray bursts (XRBs). Our aim is to understand the sensitivity of a propagating helium flame to different nuclear physics.  Using the \castro\ simulation code, we confirm the phenomenon of enhanced energy generation shortly after a flame is established after by adding ${}^{12}\mbox{C}(\mbox{p}, \gamma) {}^{13}\mbox{N}(\alpha, \mbox{p}){}^{16}\mbox{O}$ to the network, in agreement with the past literature. This sudden outburst of energy leads to a short accelerating phase, causing a drastic alteration in the overall dynamics of the flame in XRBs. 
Furthermore, we investigate the influence of different plasma screening routines on the propagation of the XRB flame. We finally examine the performance of simplified-SDC, a novel approach to hydrodynamics and reaction coupling incorporated in \castro, as an alternative to operator-splitting.

\end{abstract}


\section{Introduction}\label{Sec:Introduction}

An X-ray burst (XRB) is a thermonuclear runaway caused by the ignition of the accreted fuel on the surface of a neutron star. Through Roche-lobe overflow, accreted matter is transferred from the companion star's envelope to the surface of the neutron star. The companion star is likely to have a similar composition to the Sun, i.e.\ a mixture of hydrogen, helium, and a small fraction of carbon, nitrogen, and oxygen \citep{Galloway_2020_basics_of_xrb}. Nuclear explosions are directly affected by the initial composition of the accreted layer and the accretion rate. XRBs found with a mixture of hydrogen and helium fuel layers show bursts with $\sim 5$ sec rise time, while a pure helium layer is typically more explosive and shows bursts with shorter rise times ($\sim 1$ sec). If the shell is enriched with carbon rather than helium, then the explosion is likely to have an extended rise duration, from minutes to hours known as a superburst \citep{Kuulkers_2002,Cumming_2001,Gupta_2007}. 

Since the timescale of accretion between each burst is in the order of $10^4$ secs while the rise time is in the order of $\sim 1 - 10$ sec \citep{Parikh_2013}, it is unlikely for the same thermodynamic condition to exist on the entire surface of the neutron star and for the entire surface to start burning simultaneously \citep{Shara_1982}. Therefore, nuclear ignition is likely to begin in a localized region, and spread to the rest of the neutron star \citep{Spitkovsky_2002}. This asymmetrical burning leads to modulations in the observed flux during the rotation of neutron stars, which causes asymmetrical surface brightness during XRBs \citep{strohmayer_2009}. This explains burst oscillation behavior, a millisecond period variation of XRB intensity during rise time, and the double-peak light curve for non-photospheric-radius-expansion XRBs discovered and supported by observational data \citep{Altamirano_2010,Chakraborty_2014,Bhattacharyya_2006, Kaaret_2007,Smith_1997}.  


Numerous successful numerical simulations have explored the properties of neutron stars and XRBs.
One-dimensional simulations can determine the nucleosynthesis, study the rp-process, estimate burst duration, and predict the light curve of XRBs by assuming spherical symmetry \citep{Woosley_2004,Johnston_2018,Meisel_2018,Johnston_2020}. On the other hand, multi-dimensional simulations are conducted to study the behavior of lateral flame propagation and flame structure for their nontrivial contributions to the structure of the XRB light curve \citep{eiden:2020,harpole:2021,cavecchi:2013,Cavecchi_2015,Cavecchi_2016}.



In our first work, \cite{eiden:2020}, we studied the overall behavior of the burning front propagation using a 2D simulation with various approximations like artificially boosted flame speed and simple networks to minimize the computational cost, as well as high rotation rate to reduce the lateral flame length scale for greater flame confinement. In \cite{harpole:2021}, we continued our work to explore the effect of rotation rate and crust temperature on flame propagation. We discovered that flame propagation was not greatly affected by the rotation rate, which is likely due to the balance between the confinement of the flame and the enhanced nuclear reaction rates caused by the increased rotation rate. It is also demonstrated that a cooler crust temperature drives a reduced flame speed and vice versa. 
Finally, in \citet{xrb_flame_3d} we showed
that 2D axisymmetric simulations compare well to full 3D simulations in capturing the nucleosynthesis of the early flame propagation.
Continuing our work, we investigate the effects of different reaction networks, plasma screening methods, and time integration methods on the simulation of lateral flame propagation in XRBs. 

\section{Numerical Approach}\label{Sec:numerics}

We use {\castro} \citep{castro,castro_joss}, an open-source, adaptive mesh, astrophysical simulation code, to perform all the simulations discussed in this manuscript. The reaction networks,
integrators, equation of state \citep{timmes_swesty:2000}, thermal neutrino losses \citep{neutrino}, and conductivities
\cite{Timmes00} are contained in the related {\microphysics} package \citep{starkiller}. 

Our simulation setup is the same as described in \citep{eiden:2020, harpole:2021}, so here we only reiterate the essential points. The simulations are performed using a 2D r-z cylindrical geometry, assuming azimuthal symmetry. 
The hydrodynamics is evolved using an unsplit piecewise parabolic method \citep{ppm,ppmunsplit,millercolella:2002}.
We work with a corotating frame, and assume a rotational frequency $\Omega = 1000$ Hz for the neutron star.  Gravity is assumed to be constant, since the accreted layer is thin.  Our initial model is also unchanged from our previous papers.  We employ a hydrostatic initial model to represent the neutron star's initial thermodynamic conditions, with a hot model in the left part of the domain and a cool model in the right---this drives a rightward propagating flame (a spreading hotspot for our geometry).  By default, reactions are incorporated using
operator-splitting.


This paper investigates the effect of reaction networks on the flame.  We explore four different networks including a standard 13-isotope alpha chain as our reference network (see Sec.\ \ref{Sec:network}), two
different plasma screening implementations (Sec.\ \ref{Sec:screening}), and how the reactions and hydro are coupled
together (Sec.\ \ref{Sec:integration}).

\subsection{Nuclear Reaction Networks}\label{Sec:network}

During an XRB, nucleosynthesis involves more than a thousand isotopes \citep{Woosley_2004,Koike_2004}. However, due to computational constraints, most multi-dimensional simulations incorporate fewer than 20 isotopes. Thus, this paper aims to understand how the different approximations used in the nuclear physics impact on the lateral thermonuclear flame in XRBs.

\subsubsection{\tt aprox13}

In \citet{eiden:2020,harpole:2021}, we used the 13-isotope $\alpha$-chain network, {\tt aprox13} \citep{timmes_aprox13} network.  We'll
use that as the reference network for comparison here.  
The main feature of {\tt aprox13} is the $(\alpha, \mbox{p})(\mbox{p}, \gamma)$ approximation that eliminates the intermediate nuclei involved in $(\alpha, \mbox{p})(\mbox{p}, \gamma)$ by
assuming proton equilibrium in the reactive flow.
As a result, {\tt aprox13} has a total of 13 isotopes and 31 rates. Nevertheless, the $(\alpha, \mbox{p})(\mbox{p}, \gamma)$ approximation can become less accurate for temperatures $\gtrsim 2.5 \times 10^9$ K, potentially affecting the energy generation rates.  An illustration of this can be found in \cite{pynucastro2} (see Figure 6 there).




As XRBs can reach temperatures close to the aforementioned limit ($T \gtrsim 2.5\times 10^9$ K), it important to
understand if this approximation affects the flame.  Therefore, we employ three more intricate networks, built with \pynucastro\ using the latest rates from the {\sf REACLIB} \citep{reaclib} library.  These networks are described below.

\subsubsection{{\tt subch\_full} and {\tt subch\_full\_mod}}

The {\tt subch\_full} network was introduced in Appendix B of \cite{castro_simple_sdc} for simulating He burning in sub-Chandrasekhar mass white dwarfs (hence, the {\tt subch} prefix). {\tt subch\_full} network has three main modifications compared to {\tt aprox13} network. First {\tt subch\_full} explicitly includes the intermediate nuclei for the $(\alpha, \mbox{p})(\mbox{p}, \gamma)$ sequence.  Second, it has
a better representation of carbon and oxygen burning by including the endpoint nuclei of the difference branches: ${}^{12}\mbox{C}({}^{12}\mbox{C}, \mbox{p}) {}^{23}\mbox{Na} (\mbox{p}, \gamma) {}^{24}\mbox{Mg}$, ${}^{12}\mbox{C} ({}^{12}\mbox{C}, \mbox{n}){}^{23}\mbox{Mg} (\mbox{n}, \gamma) {}^{24}\mbox{Mg}$,  ${}^{16}\mbox{O} ({}^{16}\mbox{O}, \mbox{n}){}^{31}\mbox{S} (\mbox{n}, \gamma) {}^{32}\mbox{S}$, and ${}^{16}\mbox{O} ({}^{12}\mbox{C}, \mbox{n}){}^{27}\mbox{Si} (\mbox{n}, \gamma) {}^{28}\mbox{Si}$.  Since the neutron capture processes on the intermediate nuclei happen at a fast timescale, the rates
including neutrons are approximated by assuming the subsequent neutron capture is instantaneous as ${}^{12}\mbox{C} ({}^{12}\mbox{C}, \gamma) {}^{24}\mbox{Mg}$,  ${}^{16}\mbox{O} ({}^{16}\mbox{O}, \gamma) {}^{32}\mbox{S}$, and ${}^{16}\mbox{O} ({}^{12}\mbox{C}, \gamma) {}^{28}\mbox{Si}$, and the network does not include neutrons. We also note that their reverse rates are not included due to their negligible effects at $T \sim 10^9$ K.  Finally, we included the additional rates ${}^{14}\mbox{N}(\alpha,\gamma){}^{18}\mbox{F}(\alpha, \mbox{p}){}^{21}\mbox{Ne}$ and ${}^{12}\mbox{C}(\mbox{p}, \gamma) {}^{13}\mbox{N}(\alpha, \mbox{p}){}^{16}\mbox{O}$, discussed in \cite{Shen_2009, Weinberg_2006,Karakas_2008, fisker:2008} to bypass the comparatively slow ${}^{12}$C $\alpha$-capture process, ${}^{12}\mbox{C} (\alpha, \gamma) {}^{16}\mbox{O}$, when $T \gtrsim 10^9$ K. The first $\alpha$-capture process on ${}^{14}\mbox{N}$ and ${}^{18}\mbox{F}$ led to the production of protons that can be used for the $\alpha$ chain process on ${}^{12}\mbox{C}$. When $T \gtrsim 10^9$ K, ${}^{12}\mbox{C}(\mbox{p}, \gamma) {}^{13}\mbox{N}(\alpha, \mbox{p}){}^{16}\mbox{O}$ is expected to dominate the reactive flow towards the heavy $\alpha$ chain nuclei, which can lead to a completely different end-stage composition \citep{Weinberg_2006, fisker:2008}. Since these additional rates leave ${}^{21}\mbox{Ne}$ as the endpoint, ${}^{22}\mbox{Na}$ is added to connect all the nuclei. With all the modifications, {\tt subch\_full} has 28 isotopes and 107 rates. 
Figure~\ref{fig:networks} shows a visualization
of this network.

{\tt subch\_full\_mod} is identical to the {\tt subch\_full} network, except that the ${}^{12}\mbox{C}(\mbox{p}, \gamma) {}^{13}\mbox{N}(\alpha, \mbox{p}){}^{16}\mbox{O}$ and its reverse reactions are turned off. This is done to explore the significance of these two rates, as described in \cite{Shen_2009, Weinberg_2006, fisker:2008}. 

\subsubsection{\tt subch\_simple}

\begin{figure}
\centering
\plotone{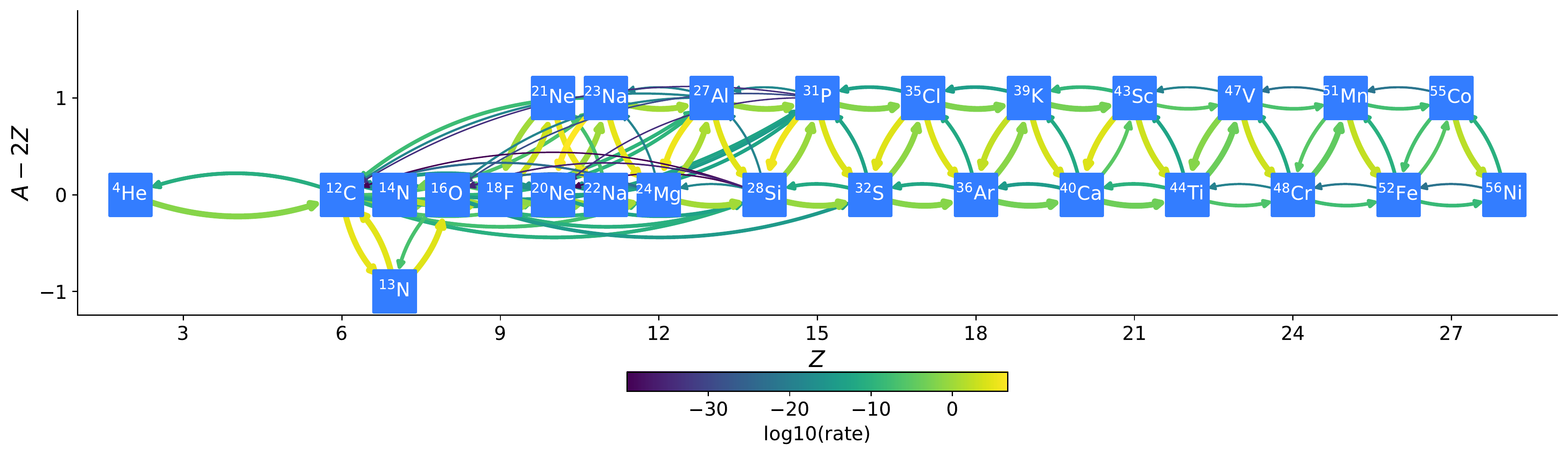}
\plotone{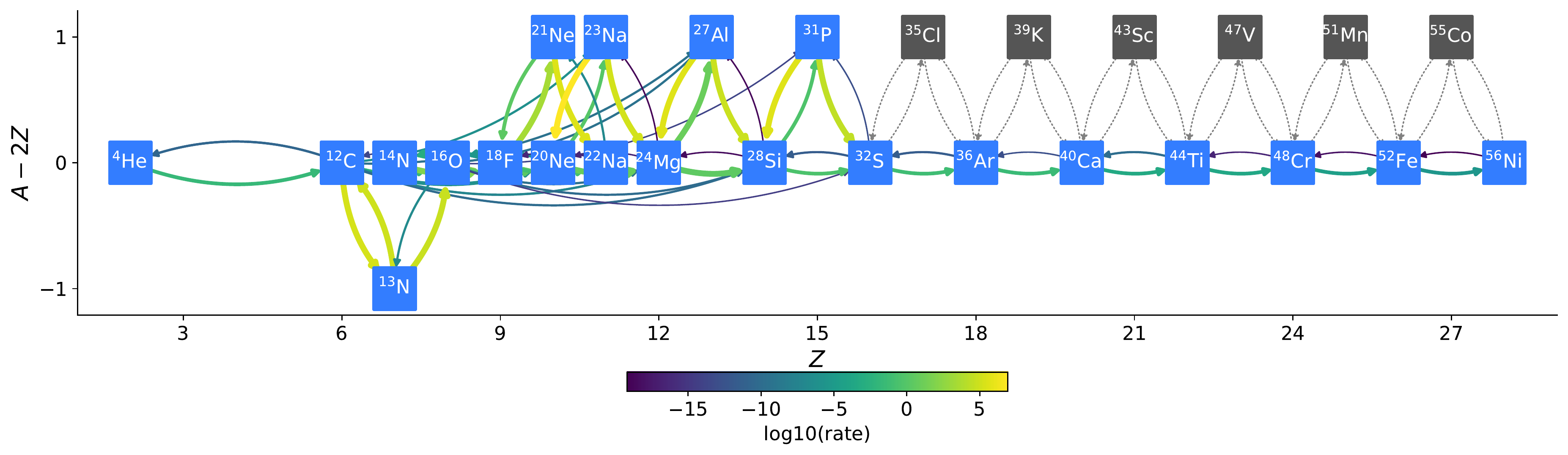}
\caption{\label{fig:networks} A simple visualization {\tt subch\_full} (top) and {\tt subch\_simple} (bottom) using the \pynucastro\ package. The color bar shows the reaction rates with solar composition, $\rho = 10^6$ g $\mathrm{cm}^{-3}$, and $T = 2 \times 10^9$ K. The horizontal axis shows the atomic number, $Z$, the vertical axis shows the extra number of neutrons compared to protons for the isotopes. The gray nodes and dotted gray lines represent the $(\alpha, \mbox{p})(\mbox{p}, \gamma)$ approximation, which are not directly in the network. All reactions that have the form, $A(X, \alpha)B$ or $A(\alpha, X)B$, are hidden for better clarity. }
\end{figure}

Lastly, we present {\tt subch\_simple} network, a simplification of {\tt subch\_full} network that would resemble a more similar network to {\tt aprox13}. The network omits a total of six nuclei, namely ${}^{35}\mbox{Cl}$, ${}^{39}\mbox{K}$, ${}^{43}\mbox{Sc}$, ${}^{47}\mbox{V}$, ${}^{51}\mbox{Mn}$, and ${}^{55}\mbox{Co}$, using the $(\alpha, \mbox{p})(\mbox{p}, \gamma)$ approximation. The reverse rates of all ${}^{12}\mbox{C} + {}^{12}\mbox{C}$,  ${}^{16}\mbox{O} + {}^{16}\mbox{O}$, and ${}^{16}\mbox{O} + {}^{12}\mbox{C}$ are removed since they are not present in {\tt aprox13}. 
All the forward and reverse rates of ${}^{12}\mbox{C} + {}^{20}\mbox{Ne}$ , ${}^{23}\mbox{Na}(\alpha, \gamma){}^{27}\mbox{Al}$, and ${}^{27}\mbox{Al}(\alpha, \gamma){}^{31}\mbox{P}$ are also removed to simplify the network.
After the simplifications, we now have 22 isotopes and 57 rates. See the bottom panel of Figure \ref{fig:networks} for visualization.

\subsection{Plasma Screening Methods}\label{Sec:screening}

Plasma screening is the enhancement of nuclear reaction rates due to the Coulomb coupling of the surrounding plasma electrons and ions. Plasma screening plays an important role in accurately calculating nuclear reaction rates in dense regions.
Depending on the thermodynamic conditions, 
screening can enhance the rate by several orders
of magnitude (see, e.g., \citealt{woosley-ignition}).
There are many approximations for screening in the literature, but these are not
often explored and compared in real simulations.
We consider two different screening approximations
for our XRB simulations.

A reaction rate cross section, $\sigma$, takes the form:
\begin{equation}
    \sigma(E) = \frac{S(E)}{E} e^{-2\pi \eta}
\end{equation}
where $S(E)$ contains the details of the nuclear physics, $E$ is the energy of particle collisions in the center-of-mass frame, and $\eta$, accounts for Coulomb barrier penetration due to quantum effects \citep{Newton2007}. The reaction rate, $R_{\textrm{th}}$, is found by integrating the cross-section over the Maxwellian velocity distribution.  If the nuclei involved in the reaction are present in a high-density region that is permeated with plasma particles, then the strength of the Coulomb barrier would be reduced due to the plasma screening effect since the effective charge of the fusing nuclei are reduced. The screening enhancement factor is usually expressed as
\begin{equation}
    F_{\textrm{scr}} = \exp{(h)}
\end{equation}
where $h$ is a function that characterizes the screening magnitude. Therefore, the overall screened nuclear reaction rate is defined as
\begin{equation}\label{Eq:screened_rate}
    R_{\textrm{scr}} = F_{\textrm{scr}}R_{\textrm{th}}
\end{equation}

The plasma screening effect exhibits varying behavior depending on the specific thermodynamic conditions. 
In the thermonuclear burning regime present in XRBs, we can divide the plasma
into weak and strong plasma screening regimes depending on the Coulomb coupling parameter of ions, $\Gamma$. A rough estimate is to consider weak plasma screening when $\Gamma \ll 1$ and strong plasma screening when $\Gamma \gtrsim 1$. In this manuscript, we explore and compare the effect of two different screening routines, {\tt SCREEN5} and {\tt CHUGUNOV2007}, available in {\microphysics} on the spread of the lateral flame propagation in XRBs.





\subsubsection{{\tt SCREEN5}}

{\tt SCREEN5} is the name of a widely-circulated screening routine (originally written in Fortran) that was the original screening routine used by \castro.
The overall procedure of {\tt SCREEN5} is summarized in the appendix of \cite{Wallace:1982}. In {\tt SCREEN5}, we define $\Gamma = (2/(Z_1+Z_2))^{1/3}Z_1 Z_2 \Gamma_e$, where $\Gamma_e = e^2\sqrt[3]{(4\pi n_e/3)}/(k_BT)$ describes the thermodynamic condition at which the reaction takes place, $Z$ is the charge of the fusing nuclei, $\mu_{12}$ is the reduced mass for the two fusing nuclei, $e$ is the electron charge, $n_e$ is the electron number density, and $k_B$ is Boltzmann's constant. When the plasma screening effect is weak or $\Gamma < 0.3$, {\tt SCREEN5} utilizes the equation proposed by \cite{Graboske_1973, Dewitt_1973}, which assumes that the interacting nuclei are separated by zero distance. To account for the spatial dependence of the screening enhancement factor, a more precise description of the strong plasma screening limit is utilized when $\Gamma > 0.8$. In \cite{jancovici:1977}, a quadratic dependence of the separation distance between the interacting nuclei was shown. This idea was applied to a one-component plasma by \cite{alastuey:1978}. By following a similar procedure outlined in \cite{itoh:1979}, the screening routine for one-component plasma can be extended to a multi-component plasma, which is suitable for a general mixture of ions. Finally, in
the intermediate screening regime, $0.3 < \Gamma < 0.8$, a weighted average between the weak and strong screening enhancement functions is used.

\subsubsection{{\tt CHUGUNOV2007}}

The overall implementation of {\tt CHUGUNOV2007} screening routine follows \cite{Chugunov_2007} with some modifications based on \cite{Yakovlev_2006} to extend calculations in a one-component plasma to multi-component plasma. The screening function, $h$, proposed in \cite{Chugunov_2007} used semi-classical calculations by assuming WKB Coulomb barrier penetration through the radial mean-field potential. Unlike {\tt SCREEN5} that uses separate expressions for different screening regimes, {\tt CHUGUNOV2007} employs a single expression that takes into account all screening limits up to $\Gamma \sim 600$. Additionally, we note that {\microphysics} includes other screening routines proposed by \cite{Chugunov_2009} and \cite{Chabrier:1998, Calder:2007}, although they are not covered in this discussion.

\subsection{Time Evolution Methods}\label{Sec:integration}

The default method for coupling hydrodynamics and nuclear reactions in {\castro} is the classic Strang-splitting method \citep{strang:1968}. This operator-splitting approach considers the hydrodynamics and nuclear reactions to be independent processes. 
The overall procedure of Strang-splitting is as follows: first, the reactive part of the system is integrated over half of the timestep, $\Delta t/2$, using a standard ODE solver to determine the solution of an intermediate state, $\mathcal{U}^\star$, which is centered in time. Advection then
evolves $\mathcal{U}^\star$ through a full timestep $\Delta t$
yielding the state $\mathcal{U}^{\star\star}$.  Finally,
the second half-timestep of burning is done for $\Delta t/2$ starting
with $\mathcal{U}^{\star\star}$ to obtain the final state $\mathcal{U}^{n+1}$. By applying the advection and hydrodynamic operator to an intermediate state that has already incorporated the nuclear reaction effect over $\Delta t/2$, an indirect coupling is achieved between the two operations and achieves a second-order accuracy in time.  The Strang-splitting implementation
in \castro\ was described in \cite{castro_strang}.

Spectral deferred correction (SDC) algorithms  \citep{dutt2000sdc, bourlioux2003reaction_sdc, castro-sdc} are iterative schemes for constructing higher-order accuracy solutions for ODEs by solving correction terms using low-order accuracy solvers like forward and backward Euler solvers. Correction terms are computed during each iteration to give improved solutions which are used as a source term in the next iteration. An arbitrarily high-order accuracy solution is then achieved after a series of correction sweeps. 


A simplfiied-SDC algorithm was introduced in \castro\ in \cite{castro_simple_sdc}.  The simplified-SDC algorithm explicitly couples reactions and hydrodynamics by including a reactive source in the hydrodynamics
interface state prediction and including an advective source in the reaction ODE integration.  
The simplified-SDC scheme offers advantages over the traditional Strang-splitting method even though they both achieve second-order accuracy in time. Specifically, the simplified-SDC scheme provides a direct coupling between hydrodynamics and nuclear reactions, which eliminates the splitting error associated with operator splitting. It also reduces the stiffness of solving reaction equations, leading to reduced computational expenses under extreme thermodynamic conditions.  In \cite{castro_simple_sdc}, it was observed that the simplified-SDC method
provides a much more accurate evolution than the Strang-split method when evolving He and C detonations
in white dwarfs.

\section{Simulations and Results}\label{Sec:results}

We present a total of seven XRB simulations using different combinations of reaction networks, time evolution methods, and screening routines. In order to investigate the sensitivity of the propagating flame to nuclear reactions, the default screening routine, {\tt SCREEN5}, and Strang-splitting are used with the four different networks: {\tt aprox13}, {\tt subch\_full}, {\tt subch\_full\_mod}, {\tt subch\_simple}. We replaced the {\tt SCREEN5} screening routine with {\tt CHUGUNOV2007} for the {\tt aprox13} model with Strang-splitting to investigate the difference in their performance. Lastly, in order to test the performance of the simplified SDC method, we ran two additional simulations using {\tt aprox13} and {\tt subch\_full} with simplified SDC instead of Strang-splitting. The overall summary of the simulations is shown in Table \ref{Tab:settings}.  We will
refer to the simulations by the names in the table
in the following discussion.

\begin{table*}
\caption{\label{Tab:settings}
Various settings used for each simulation.
}
\begin{ruledtabular}
\begin{tabular}{cccc}
Name &
Network &
Integration &
Screening
\\ 

\colrule

{\tt aprox13} & {\tt aprox13} & Strang-splitting & {\tt SCREEN5} \\
{\tt subch\_full} & {\tt subch\_full} & Strang-splitting & {\tt SCREEN5} \\
{\tt subch\_full\_mod} & {\tt subch\_full\_mod} & Strang-splitting & {\tt SCREEN5} \\
{\tt subch\_simple} & {\tt subch\_simple} & Strang-splitting & {\tt SCREEN5} \\
{\tt aprox13\_sdc} & {\tt aprox13} & simplified SDC & {\tt SCREEN5} \\
{\tt subch\_full\_sdc} & {\tt subch\_full} & simplified SDC & {\tt SCREEN5} \\
{\tt aprox13\_chu} & {\tt aprox13} & Strang-splitting & {\tt CHUGUNOV2007} 

\end{tabular}
\end{ruledtabular}
\end{table*}

\subsection{Reaction Network Comparison}\label{Sec:result_network}


\begin{figure*}
\centering
\plotone{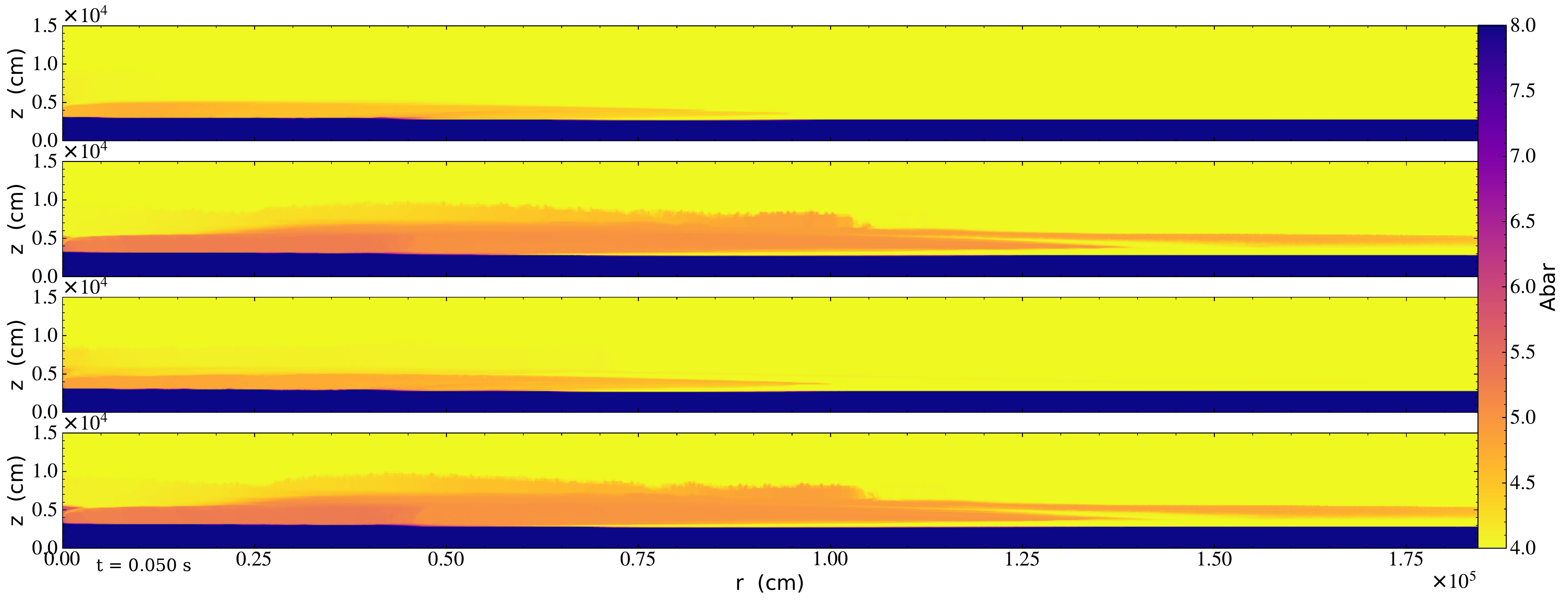}
\caption{\label{fig:network_abar_50ms} Slice plots comparing $\Bar{A}$ for {\tt aprox13} (top panel), {\tt subch\_full} (second panel from top), {\tt subch\_full\_mod} (third panel), and {\tt subch\_simple} (last panel) at 50 ms. }
\end{figure*}

\begin{figure*}
\centering
\plotone{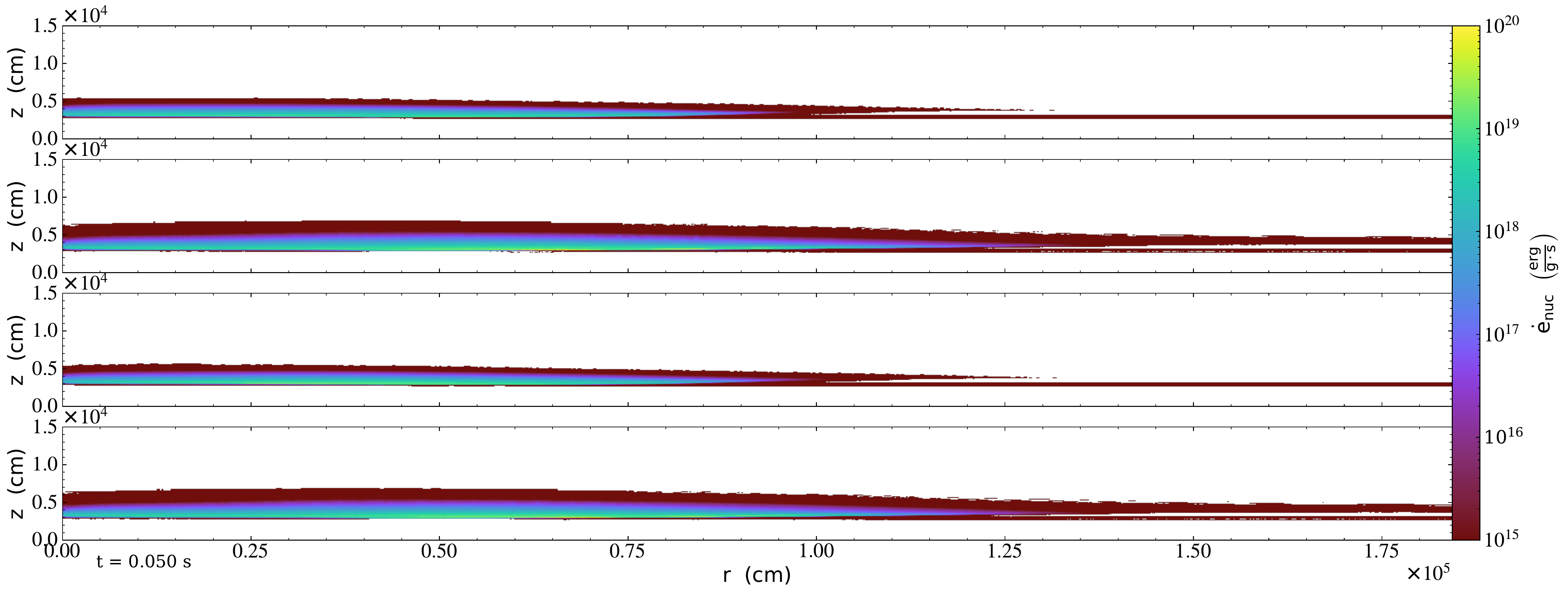}
\caption{\label{fig:network_enuc_50ms} Slice plots comparing $\dot{e}_{\textrm{nuc}}$ for {\tt aprox13} (top panel), {\tt subch\_full} (second panel from top), {\tt subch\_full\_mod} (third panel), and {\tt subch\_simple} (last panel) at 50 ms.}
\end{figure*}


\subsubsection{Global behavior}

Figure \ref{fig:network_abar_50ms} shows a comparison of the mean molecular weight, $\Bar{A}$, for the four networks at $t = 50$ ms. Regions with a larger $\Bar{A}$ represent the ash, tracing
the burning in the accreted layer. One immediate feature is the similarity of the ash structure between {\tt aprox13} and {\tt subch\_full\_mod}, as well as between {\tt subch\_full} and {\tt subch\_simple}. For {\tt subch\_full} and {\tt subch\_simple}, there is a thicker overall ash structure on the surface indicating much more vigorous burning. The sudden increase in the ash height from $r = 5 \times 10^4$ cm to $10^5$ cm for these two simulations further implies a non-uniform burning has taken place, unlike in {\tt aprox13} and {\tt subch\_full\_mod}. The darker color means that the ash in {\tt subch\_full} and {\tt subch\_simple} is composed of heavier nuclei suggesting a faster reactive flow burning to the heavier nuclei compared to the other two models. The ash also extends further out suggesting a faster flame speed compared to {\tt aprox13} and {\tt subch\_full\_mod}. 
The energy generation rate (shown in Figure \ref{fig:network_enuc_50ms}) shows the same trends: {\tt subch\_full} and {\tt subch\_simple} clearly possess a higher $\dot{e}_{\textrm{nuc}}$ in both magnitude and region coverage compared to the other two. 

\begin{figure*}
\centering
\plotone{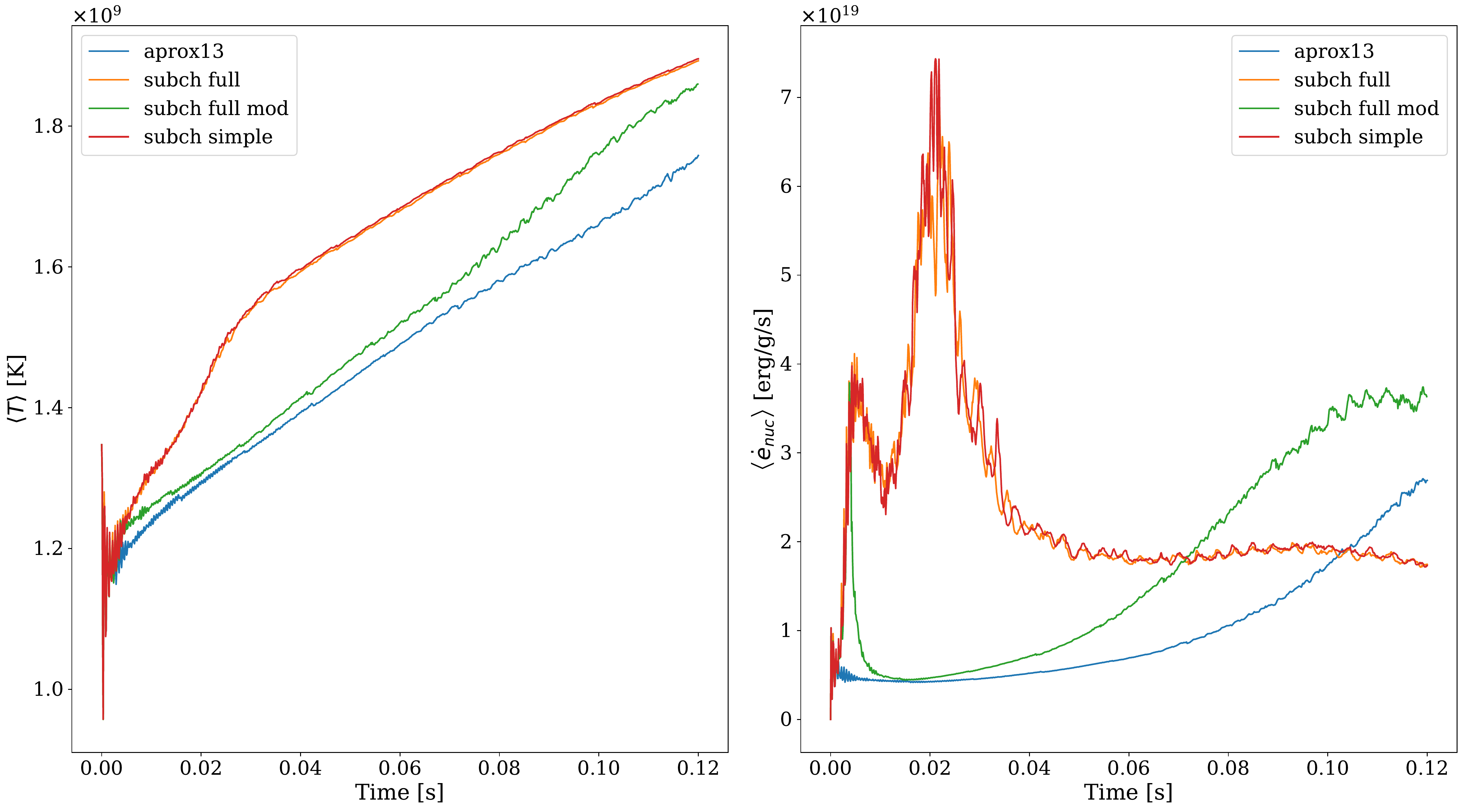}
\caption{\label{fig:network_time_profile} Time profiles showing the weighted temperature (left panel) and energy generation rate (right panel) of the burning front for the 4 simulations models: {\tt aprox13}, {\tt subch\_full}, {\tt subch\_full\_mod}, and {\tt subch\_simple}.}
\end{figure*}
 
In order to quantify how the flame changes over time, we look at the
density-weighted profile of temperature and energy generation rate in the burning, following the procedure in \citet{harpole:2021}. 
We selected cells that are in the 99th percentile or higher for temperature or energy generation rate and compute a density-weighted average of temperature and energy generation rate:
\begin{equation}\label{Eq:weighted_equation}
    \left<Q\right>_w = \frac{\sum\limits_{c_i}^{} \rho(c_i) Q(c_i)}{\sum\limits_{c_i} \rho(c_i)}; c_i \in C_{99}(Q)
\end{equation}
where $\left<Q\right>_w$ is the weighted quantity, $C_{99}(Q)$ are cells where the quantity, $Q$, is in the 99th percentile or higher, and $\rho(c_i)$ and $Q(c_i)$ are the density and the quantity in $c_i$ cell.
Figure \ref{fig:network_time_profile} shows the resulting profiles. 

The temperature and energy generation rate profiles for {\tt subch\_full} and {\tt subch\_simple} are nearly identical.  As the primary distinction between these two is the $(\alpha, \mbox{p})(\mbox{p}, \gamma)$ approximation for heavier nuclei, we can conclude that this approximation remains accurate in the context of XRB. 

In general, there is no uniform shape in both T and $\dot{e}_{\textrm{nuc}}$ for {\tt subch\_full} and {\tt subch\_simple}. After the flame-establishing phase at $\sim 3$ ms, there is a slight acceleration in temperature followed by a small decline, corresponding to the first spike in $\dot{e}_{\textrm{nuc}}$. However, there is a sudden burst of energy production output from $\sim 10$ ms after the first spike for {\tt subch\_full} and {\tt subch\_simple}. The sudden spike in energy production in {\tt subch\_full}, as compared to {\tt subch\_full\_mod}, is attributed to the inclusion of ${}^{12}\mbox{C}(\mbox{p}, \gamma) {}^{13}\mbox{N}(\alpha, \mbox{p}){}^{16}\mbox{O}$ rates, as it is the only distinction between these two networks. By looking at the weighted $\dot{e}_{\textrm{nuc}}$ profile, we determined that the burning acceleration phase had ended before 50 ms, as anticipated from previous slice plots. In general, the changes in $\dot{e}_{\textrm{nuc}}$ are well reflected on the weighted-temperature profile. 

Unlike {\tt subch\_full} or {\tt subch\_simple}, {\tt aprox13} and {\tt subch\_full\_mod} have a much more steady burning. Initially, {\tt subch\_full\_mod} appears to keep pace with {\tt subch\_full} and {\tt subch\_simple}, but $\dot{e}_{\textrm{nuc}}$ declines rapidly after reaching its peak following the flame-establishing phase. This is the consequence of the missing ${}^{12}\mbox{C}(\mbox{p}, \gamma) {}^{13}\mbox{N}(\alpha, \mbox{p}){}^{16}\mbox{O}$, which puts a heavy limitation on the subsequent $\alpha$-chain burning processes toward the heavy elements. However, $\dot{e}_{\textrm{nuc}}$ for {\tt subch\_full\_mod} gradually accelerates, reaching its peak at 120 ms similar to {\tt aprox13}, but with a faster rate. In contrast, $\dot{e}_{\textrm{nuc}}$ for {\tt subch\_full} and {\tt subch\_simple} gradually decrease during the later stages of burning after the burst.

\begin{figure*}
    \centering
    \plotone{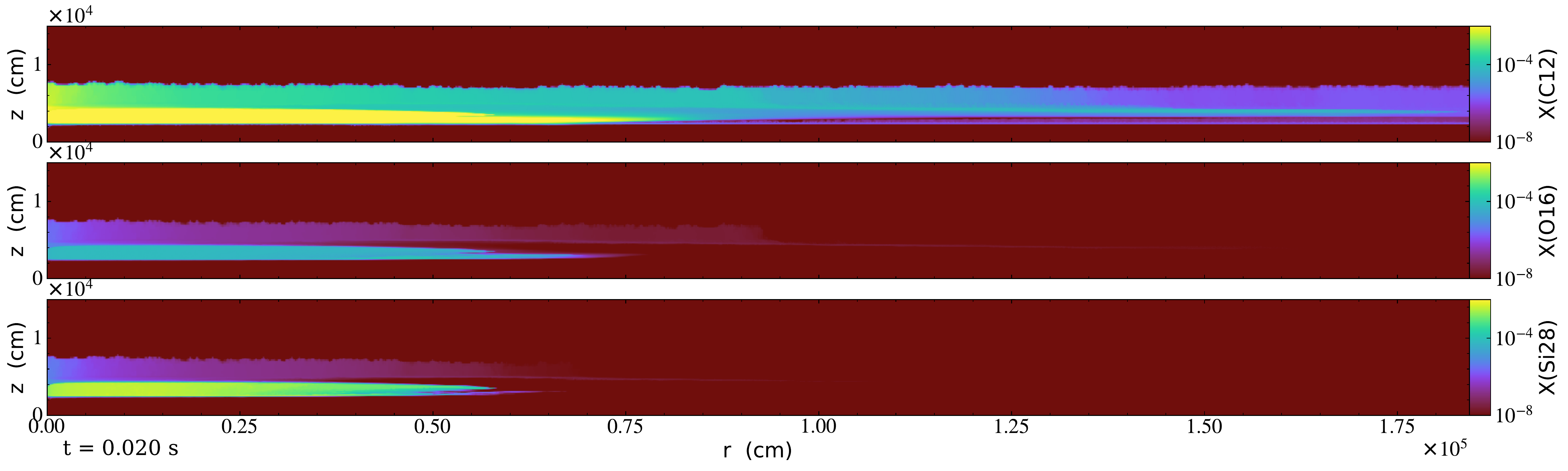}
    \caption{\label{fig:network_aprox13_nuc} Slice plots showing the mass fractions of ${}^{12}$C, ${}^{16}$O, and ${}^{28}$Si for {\tt aprox13} at 20 ms.}
\end{figure*}

\begin{figure*}
    \centering
    \plotone{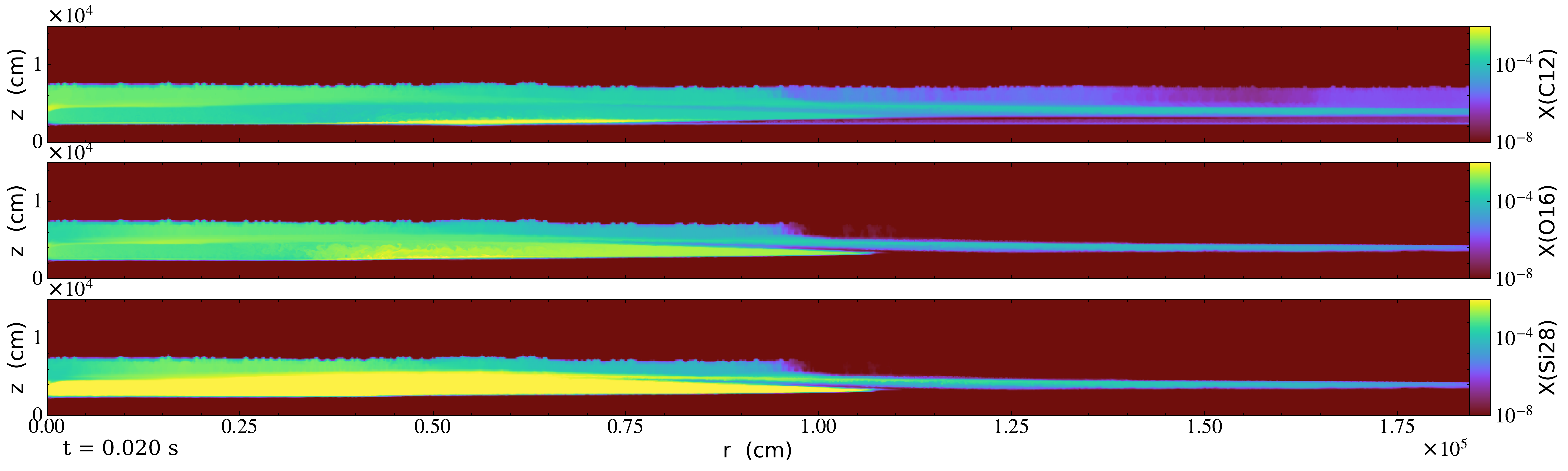}
    \caption{\label{fig:network_subch_nuc} Slice plots showing the mass fractions of ${}^{12}$C, ${}^{16}$O, and ${}^{28}$Si for {\tt subch\_full} at 20 ms.}
\end{figure*}

\begin{figure*}
    \centering
    \plotone{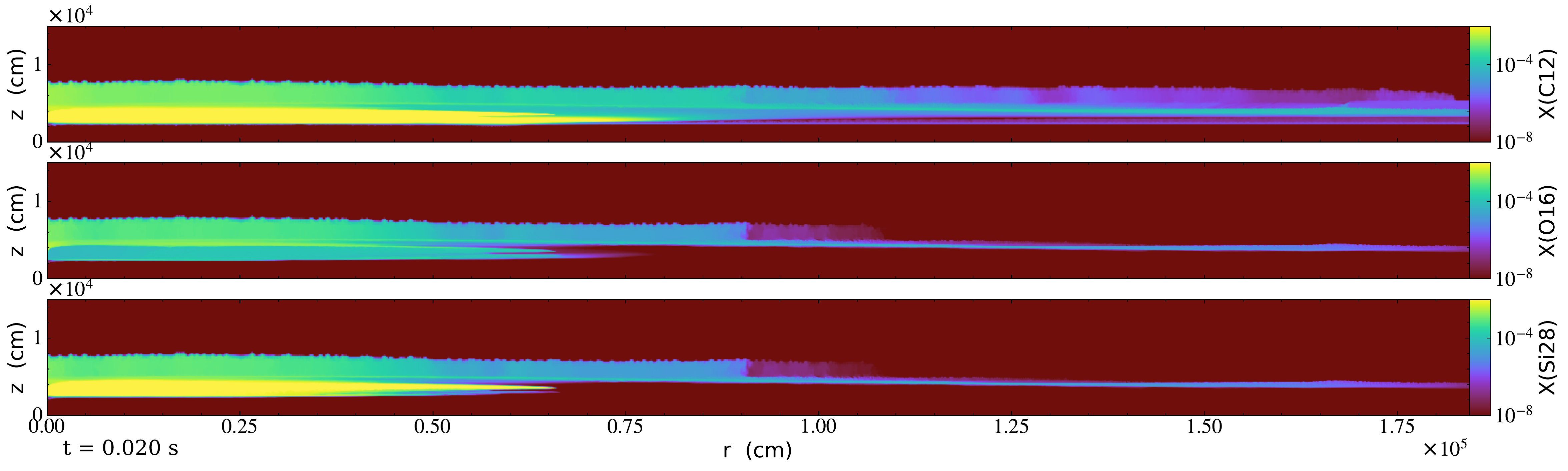}
    \caption{\label{fig:network_subch_mod_nuc} Slice plots showing the mass fractions of ${}^{12}$C, ${}^{16}$O, and ${}^{28}$Si for {\tt subch\_full\_mod} at 20 ms.}
\end{figure*}

\subsubsection{Nucleosynthesis}

In order to investigate the burning process in detail, Figure \ref{fig:network_aprox13_nuc}, \ref{fig:network_subch_nuc}, and \ref{fig:network_subch_mod_nuc} show the mass fractions of ${}^{12}$C, ${}^{16}$O, and ${}^{28}$Si for {\tt aprox13}, {\tt subch\_full}, and {\tt subch\_full\_mod} at 20 ms, respectively. We exclude {\tt subch\_simple} due to its similarity to {\tt subch\_full}.  For {\tt aprox13}, all the burning products are concentrated in a thin region in the left of the domain, as we would expect
given the high temperature behind the flame.  Notably, there is an abundance of ${}^{12}$C, as {\tt aprox13} relies on the relatively slow $\alpha$-capture rate to convert ${}^{12}$C to ${}^{16}$O. Additionally, ${}^{16}$O is predominately transformed into ${}^{28}$Si, which is the primary end product of the network.

With {\tt subch\_full}, we see that the ${}^{12}$C is much more depleted in the burning region and there is much more production of heavy $\alpha$ chain isotopes through the $\alpha$-chain, particularly ${}^{28}$Si, at $t = 20$ ms. This phenomenon is consistent with the results of \cite{Weinberg_2006, fisker:2008}. The burning path, ${}^{12}\mbox{C}(\mbox{p}, \gamma) {}^{13}\mbox{N}(\alpha, \mbox{p}){}^{16}\mbox{O}$, provides an efficient route for converting ${}^{12}$C to ${}^{16}$O compared to the $\alpha$-capture rate. 

Similar to {\tt aprox13}, {\tt subch\_full\_mod} exhibits a considerable amount of ${}^{12}$C due to the absence of ${}^{12}\mbox{C}(\mbox{p}, \gamma) {}^{13}\mbox{N}(\alpha, \mbox{p}){}^{16}\mbox{O}$ rate. However, unlike {\tt aprox13}, a higher concentration of ${}^{16}$O is observed in the upper vertical regions in {\tt subch\_full\_mod}. This is likely the result of incorporating various additional rates, including more up-to-date rates from {\sf REACLIB} library, in {\tt subch\_full\_mod} compared to {\tt aprox13}. Nevertheless, the quantity of ${}^{16}$O in the bottom regions remains similar to those seen in {\tt aprox13}, as the slightly higher temperature in the bottom region facilitates the burning of these fuels into ${}^{28}$Si.

\begin{figure*}
    \centering
    \plotone{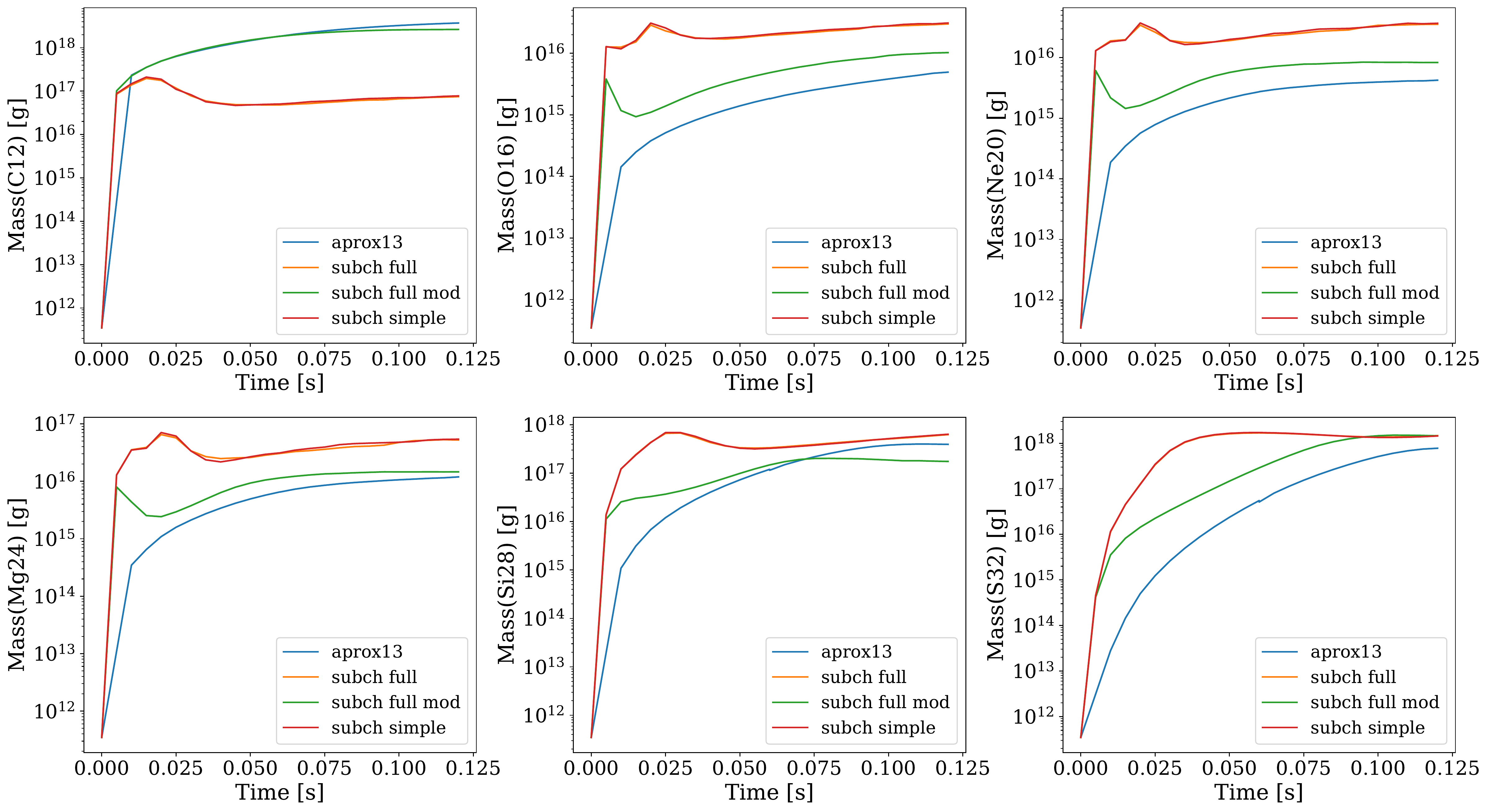}
    \caption{\label{fig:network_species_summary} The overall evolution of the total mass for ${}^{12}$C, ${}^{16}$O, ${}^{20}$Ne, ${}^{24}$Mg, ${}^{28}$Si, and ${}^{32}$S for the 4 simulations models: {\tt aprox13}, {\tt subch\_full}, {\tt subch\_full\_mod}, and {\tt subch\_simple}.}
\end{figure*}

Figure \ref{fig:network_species_summary} shows the total mass of the key $\alpha$-chain isotopes to track their evolution. The initial spike in the production of ${}^{16}$O, ${}^{20}$Ne, ${}^{24}$Mg, corresponds to the initial peak in energy generation rates for all {\tt subch} networks shown in Figure \ref{fig:network_time_profile}. Although ${}^{12}\mbox{C}(\mbox{p}, \gamma) {}^{13}\mbox{N}(\alpha, \mbox{p}){}^{16}\mbox{O}$ efficiently burns ${}^{12}$C into ${}^{16}$O, resulting in a lower abundance of ${}^{12}$C in {\tt subch\_full} and {\tt subch\_simple}, a considerable amount of ${}^{12}$C still accumulates before $t \sim 18$ ms. Nevertheless, when $t \gtrsim 18$ms, corresponding to $T \sim 1.3 \times 10^9$ K, the flow from ${}^{12}$C to ${}^{16}$O can surpass the triple-$\alpha$ process, resulting in a depletion of ${}^{12}$C. This phenomenon is consistent with results in \cite{Weinberg2007, fisker:2008}. Consequently, there is an amplification in the nuclear energy generation rates and the mass production of heavy isotopes such as ${}^{28}$Si. The early exhaustion of ${}^{12}$C also causes a shortage of burning fuels during the subsequent burning stage for {\tt subch\_full} and {\tt subch\_simple}. This is because the nuclear burning process is now bottle-necked by the triple-$\alpha$ process, corresponding to an overall decline in the energy generation rate following the outburst. In contrast, both {\tt aprox13} and {\tt subch\_full\_mod} exhibit a continuous accumulation of ${}^{12}$C since the networks are bottle-necked by the inefficient $\alpha$-capture rate on ${}^{12}$C to ${}^{16}$O. However, unlike {\tt aprox13}, {\tt subch\_full\_mod} demonstrates slightly more effective burning paths for burning ${}^{12}$C at $t \gtrsim 70$ ms. This phenomenon likely accelerates the overall nuclear burning process, resulting in significantly higher nuclear energy generation compared to {\tt aprox13} during the late-stage burning, as shown in Figure \ref{fig:network_time_profile}.

\begin{figure*}
    \centering
    \plotone{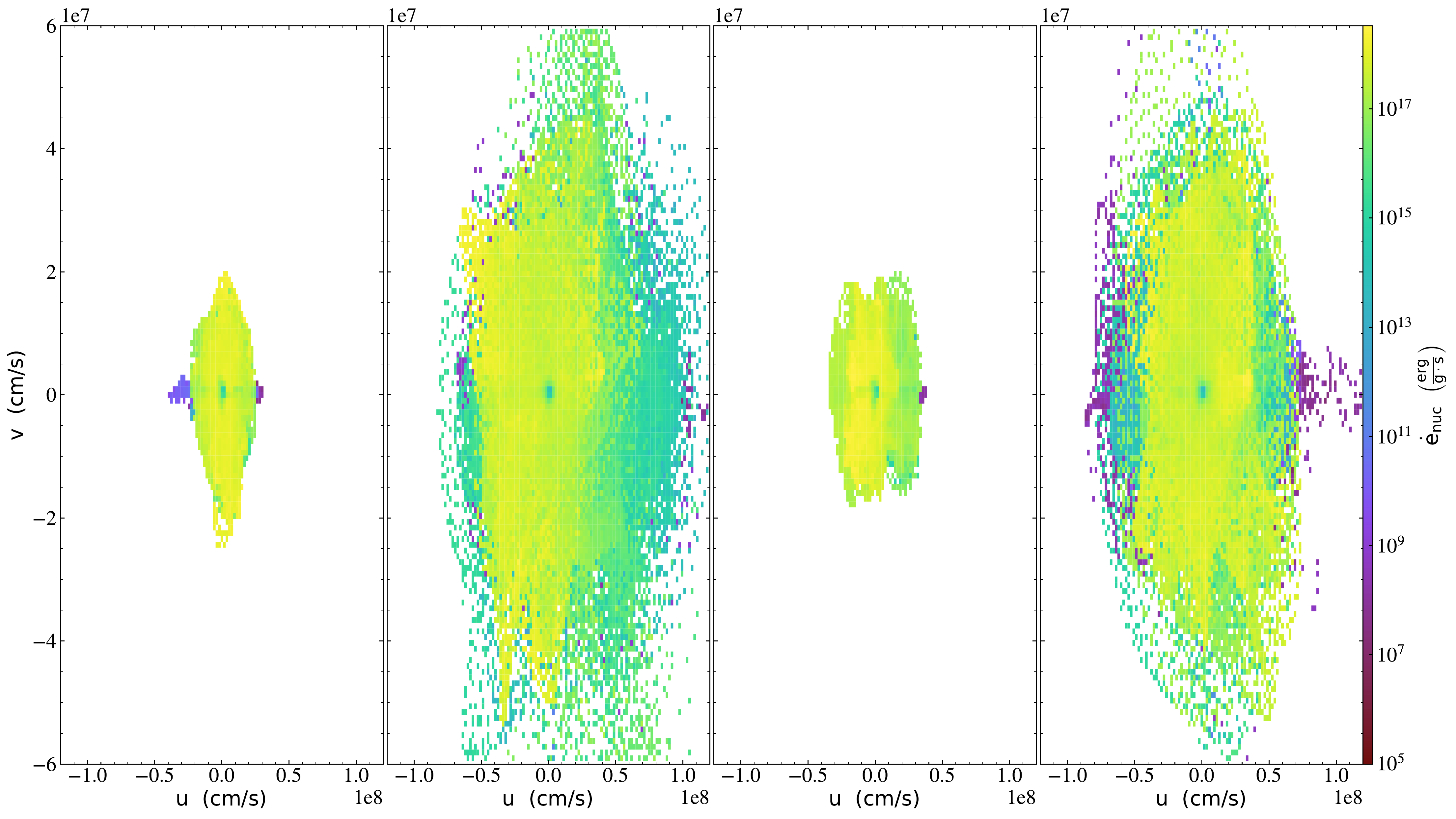}
    \plotone{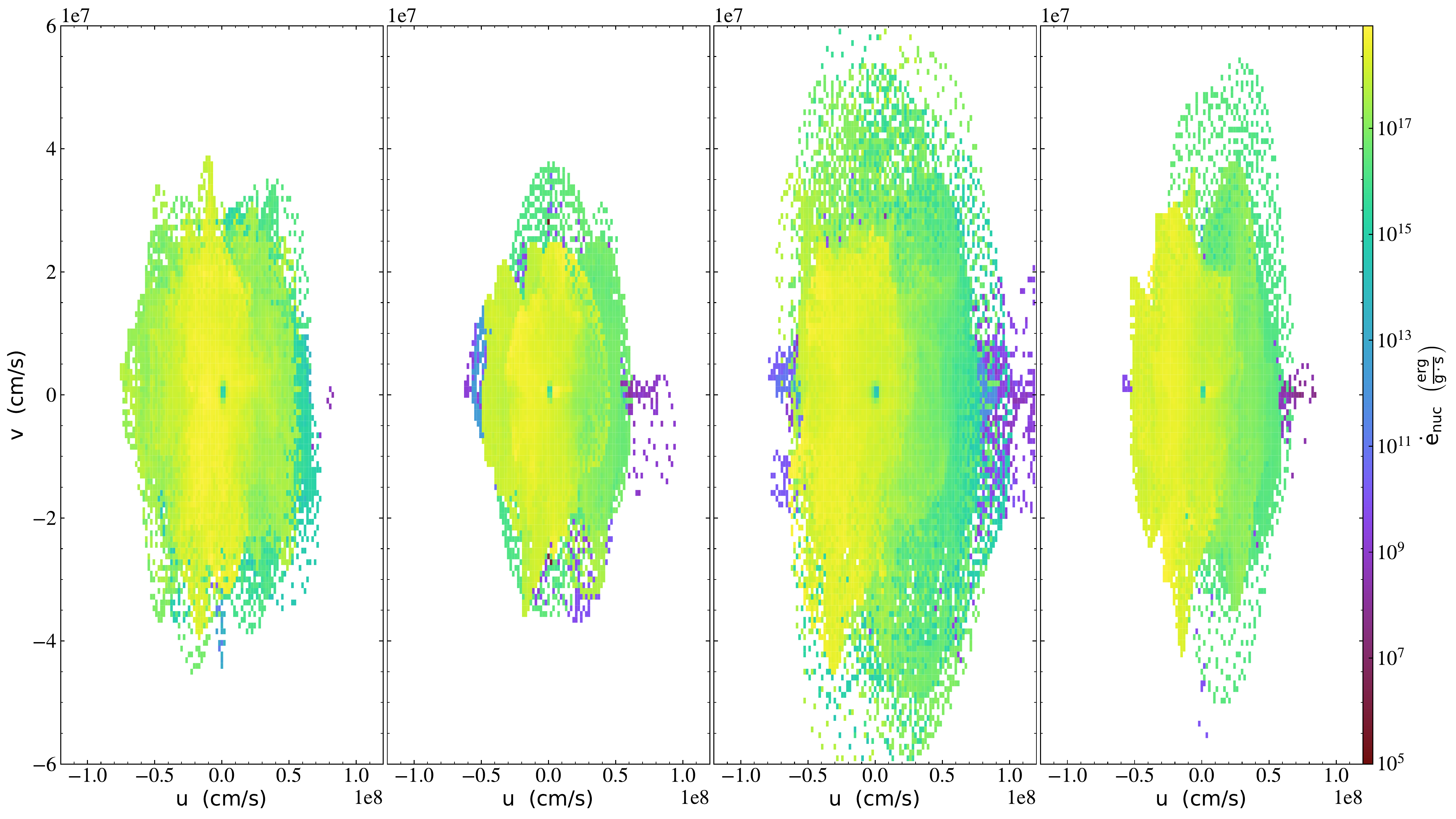}
    \caption{\label{fig:network_u_v_enuc} $u-v$ phase plots for {\tt aprox13} (panels in the first column on the left), {\tt subch\_full} (second column), {\tt subch\_full\_mod} (third column), {\tt subch\_simple} (fourth column) at $t = 25$ ms (top 4 panels) and $t = 100$ ms (bottom 4 panels). The x-axis, $u$, shows the velocity in the $r$ direction, whereas the y-axis, $v$, shows the velocity in the $z$ direction. The color bar shows $\dot{e}_{\textrm{nuc}}$.}
\end{figure*}

\subsubsection{Dynamics}

Figure \ref{fig:network_u_v_enuc} shows the $u$-$v$ phase plot (radial velocity vs.\ vertical velocity) for the four simulations models at $t = 25$ ms and $t = 100$ ms, colored by $\dot{e}_{\textrm{nuc}}$. Notably, the peak of the energy spike occurs at $t \sim 25$ ms for {\tt subch\_full} and {\tt subch\_simple},  resulting in the comparatively larger distribution in their $u$-$v$ phase space plots.  By 100~ms, the distribution appears to shrink considerably, likely due to decrease in the energy output as the flame is
established. In contrast, the $u$-$v$ phase space distribution for {\tt aprox13} and {\tt subch\_full\_mod} increased from $t = 25$ ms to $t = 100$ ms. However, there is one common feature slowly forming in the late stage where higher $\dot{e}_{\textrm{nuc}}$ are preferably located in regions with a velocity opposite to the flame propagation. This a common feature observed in \cite{eiden:2020, harpole:2021}.

\begin{figure*}
\centering
\plotone{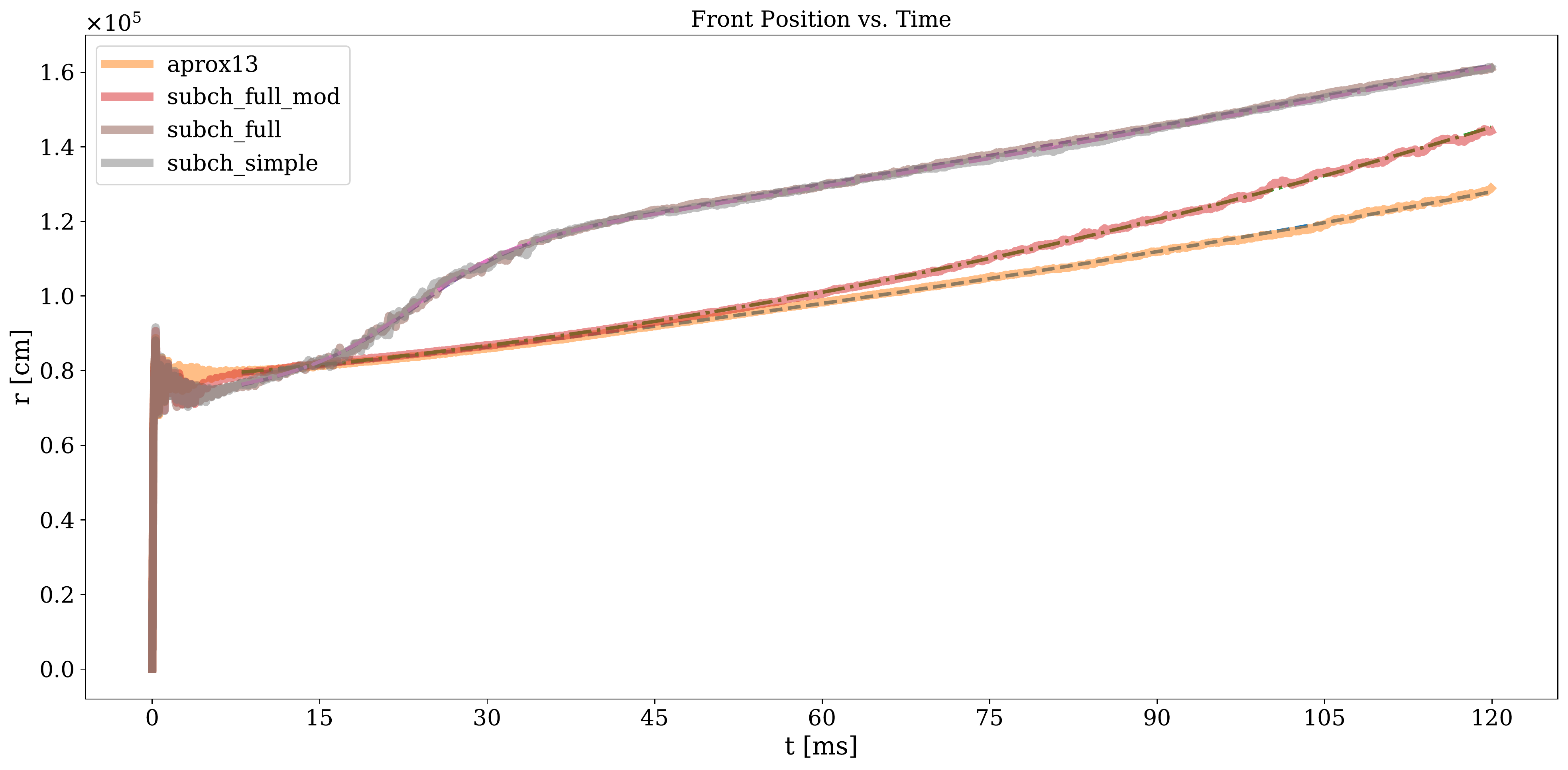}
\caption{\label{fig:network_front} Flame front position as a function time for {\tt aprox13}, {\tt subch\_full}, {\tt subch\_full\_mod}, and {\tt subch\_simple}. The results of the fitting function, Eq. \ref{Eq:quadratic_fit} and \ref{Eq:tanh_fit}, is also shown in the dashed lines.}
\end{figure*}

To examine the overall impact of different reaction networks on the dynamics of the laterally propagating flame, Figure \ref{fig:network_front} shows the radial flame front position as a function of time. We follow the approach outlined in \cite{eiden:2020}, where the position of the flame front is defined as the location at which $\dot{e}_{\textrm{nuc}}$ of the 1D radial profile first drops to $0.1\%$ of the global maximum $\dot{e}_{\textrm{nuc}}$ for $r > r_{\mbox{max}(\dot{e}_{nuc})}$. As expected, {\tt subch\_full} and {\tt subch\_simple} have nearly identical burning front positions, with a sudden burst of acceleration followed by a gradual decline in velocity. These behaviors are consistent with the temperature and $\dot{e}_{\textrm{nuc}}$ profiles we examined previously. On the other hand, {\tt subch\_full\_mod} and {\tt aprox13} exhibit a similar gradually accelerating burning front position, with a faster acceleration for {\tt subch\_full\_mod}.

\begin{table*}
\caption{\label{Tab:network_fitted_param}
Fitted parameters of the fitting functions Eq. \ref{Eq:quadratic_fit} and \ref{Eq:tanh_fit} for {\tt aprox13}, {\tt subch\_full}, {\tt subch\_full\_mod}, and {\tt aprox13\_chu}. The fitting function is applied for $t > 8$ ms.
}

\begin{ruledtabular}
\footnotesize
\centering
\begin{tabular}{ccccccc}
\small 
Name &
$a_0$ [$\mbox{km} \ \mbox{s}^{-2}$] &
$v_0$ [$\mbox{km} \ \mbox{s}^{-1}$] &
$r_0$ [km] &
A [km]&
B [s]&
C

\\ 
\colrule
{\tt aprox13} & $24.22 \pm 0.23$ & $2.812 \pm 0.015$ & $0.7680\pm 0.0004 $ & N/A & N/A & N/A\\

{\tt subch\_full} & $9.22 \pm 0.82$ & $4.489 \pm 0.067$ & $0.8646 \pm 0.0014$ & $0.150 \pm 0.001$ & $0.0093 \pm 0.0001$ & $-2.551 \pm 0.035$ \\

{\tt subch\_full\_mod} & $58.53 \pm 0.24$ & $2.122 \pm 0.016$ & $0.7773 \pm 0.0004$ & N/A & N/A & N/A \\

{\tt subch\_simple} & $15.60 \pm 0.93$ & $3.961 \pm 0.076$ & $0.8745 \pm 0.0016$ & $0.153 \pm 0.002$ & $0.0091 \pm 0.0001$ & $-2.575 \pm 0.040$\\
 
{\tt aprox13\_chu} & $22.96 \pm 0.20$ & $2.578 \pm 0.0133$ & $0.7728 \pm 0.0004$ & N/A & N/A & N/A\\
 


\end{tabular}
\end{ruledtabular}
\end{table*}

We fit the flame front position with a simple function to estimate
the flame speed.  For the {\tt aprox13} and {\tt subch\_full\_mod} networks, we use the same expression, Eq. \ref{Eq:quadratic_fit}, 
as in \citet{harpole:2021}.
\begin{equation}\label{Eq:quadratic_fit}
    r(t) = \frac{1}{2}a_0 t^2 + v_0 t + r_0
\end{equation}
On the other hand, Eq. \ref{Eq:tanh_fit} is used for {\tt subch\_full} and {\tt subch\_simple} networks. A hyperbolic tangent function is added to the fitting function to account for the burst of acceleration at $10 \ \mbox{ms} \lesssim t \lesssim 25 \ \mbox{ms}$.
\begin{equation}\label{Eq:tanh_fit}
    r(t) = A\tanh{\left(\frac{t}{B} + C\right)} + \frac{1}{2}a_0 t^2 + v_0 t + r_0
\end{equation}
Both fitting functions are applied for $t > 8$ ms, and the fitted parameters along with their respective errors are shown in Table \ref{Tab:network_fitted_param}. The errors are calculated by taking the square root of the diagonal of the covariance matrix.

\begin{table*}
\caption{\label{Tab:network_instan_vel}
Instantaneous flame propagation speed at $t = 23$ ms and $t = 100$ ms for {\tt aprox13}, {\tt subch\_full}, {\tt subch\_full\_mod}, {\tt subch\_simple}, and {\tt aprox13\_chu}. $t = 23$ ms and $t=100$ ms represent the acceleration phase for {\tt subch\_full} and {\tt subch\_simple} and the steady phase at the late-stage, respectively. $t_{10}$ represents the theoretical time for the flame to reach 10 km.
}

\begin{ruledtabular}
\footnotesize
\centering
\begin{tabular}{cccc}
\small 
Name &
$v_{23}$ [$\mbox{km} \ \mbox{s}^{-1}$] &
$v_{100}$ [$\mbox{km} \ \mbox{s}^{-1}$] &
$t_{10}$ [s]
\\
\colrule

{\tt aprox13} & $3.369 \pm 0.016$ & $5.234 \pm 0.027$ & 0.7647 \\
{\tt subch\_full} & $20.732 \pm 0.284$ & $5.411 \pm 0.105$ & 0.9917 \\
{\tt subch\_full\_mod} & $3.468 \pm 0.017$ & $7.975 \pm 0.029$ & 0.4873 \\
{\tt subch\_simple} & $21.095 \pm 0.332$ & $5.521 \pm 0.120$ & 0.8483 \\
{\tt aprox13\_chu} & $3.106 \pm 0.014$ & $4.874 \pm 0.024$ & 0.7912 \\
\end{tabular}
\end{ruledtabular}
\end{table*}

Using the fitted parameters, the instantaneous speed of the flame front at different times can be calculated, as shown in Table \ref{Tab:network_instan_vel}. We observe that during the acceleration burst phase at $t = 23$ ms, the difference in the instantaneous speed between {\tt subch\_full} and {\tt subch\_simple} can be as much as 7 times higher compared to the other two simulations. Furthermore, {\tt subch\_full\_mod} accelerates towards the end since it still has a sufficient amount of ${}^{12}$C fuel due to the inefficient reaction flows to ${}^{16}$O. Even though the flame front for {\tt subch\_full\_mod} is expected to surpass {\tt subch\_full} and {\tt subch\_simple} at $t \sim 160$ ms, based on Figure \ref{fig:network_front}, $\dot{e}_{\textrm{nuc}}$ from Figure \ref{fig:network_time_profile} appears to reach its peak at $120$ ms. Therefore, there is no guarantee that the lateral propagating flame for {\tt subch\_full\_mod} would exceed its counterparts. 

Assuming that the fitting functions accurately describe the flame propagation, the expected time for the flame to reach a typical neutron star radius of 10 km, $t_{10}$, can be calculated, as shown in Table \ref{Tab:network_instan_vel}. $t_{10}$ serves as a rough prediction for the rise time of XRBs with a pure ${}^{4}$He accretion layer. $t_{10}$ for all models are $\lesssim 1$ sec, whereas $t_{10}$ for {\tt subch\_full} is $\sim 1$ sec, consistent with previous observational studies \citep{galloway:2008}.

\subsection{Plasma Screening Routine Comparison}\label{Sec:result_screening}



\begin{figure*}
\centering
\plotone{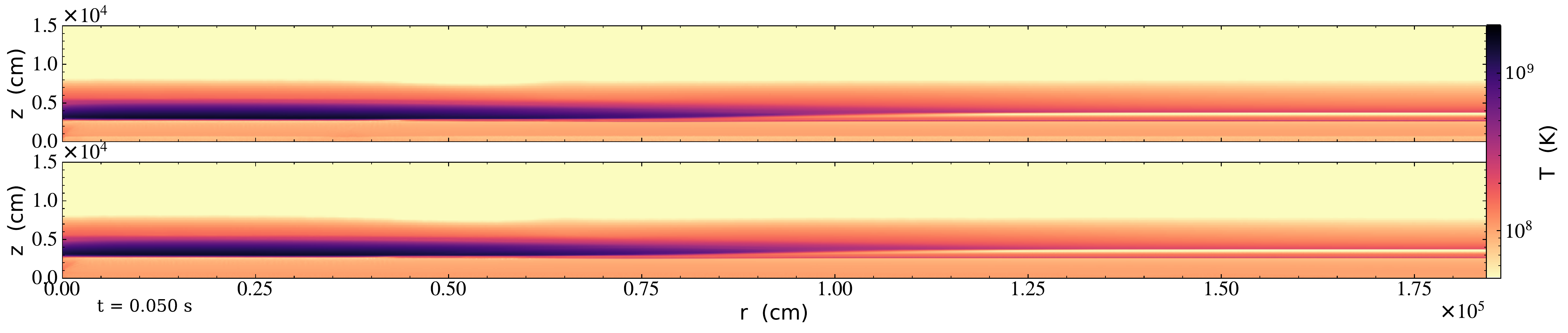}
\caption{\label{fig:screen_temp} Slice plots comparing temperature for {\tt aprox13} (top panel) and {\tt aprox13\_chu} (bottom panel) at $t = 50$ ms.}
\end{figure*}

\begin{figure*}
\centering
\plotone{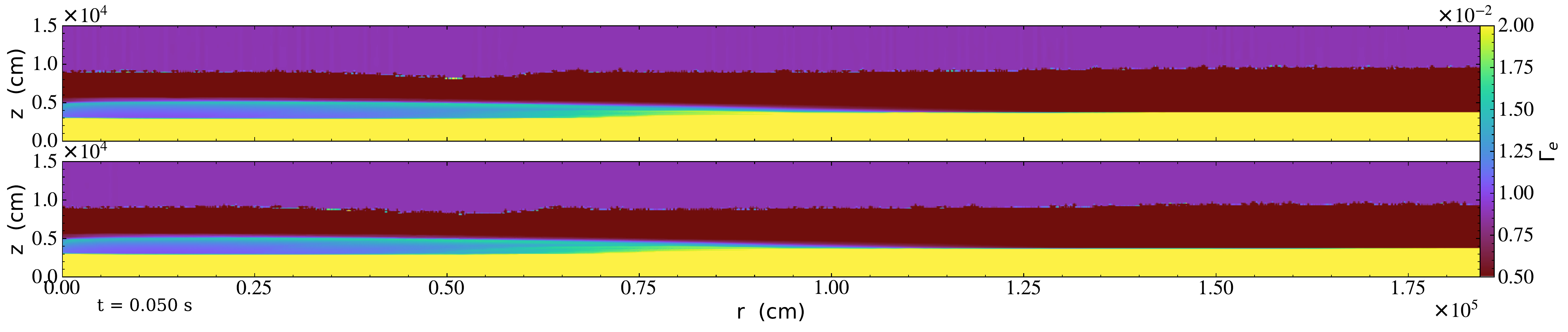}
\caption{\label{fig:screen_gamma_50ms} Slice plots comparing $\Gamma_e$ for {\tt aprox13} (top panel) and {\tt aprox13\_chu} (bottom panel) at $t = 50$ ms.}
\end{figure*}

\begin{figure*}
\centering
\plotone{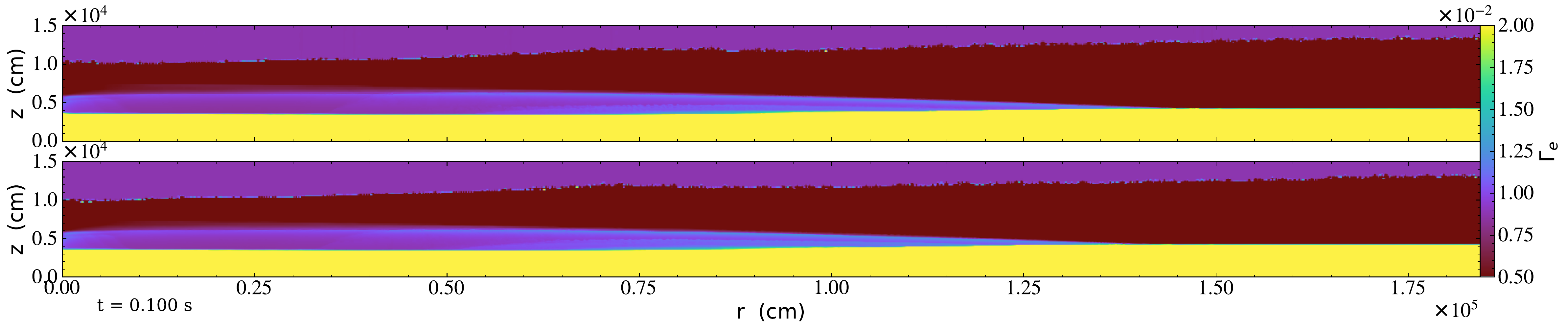}
\caption{\label{fig:screen_gamma_100ms} Slice plots comparing $\Gamma_e$ for {\tt aprox13} (top panel) and {\tt aprox13\_chu} (bottom panel) at $t = 100$ ms.}
\end{figure*}

We now present a comparison between the effects of {\tt SCREEN5} and {\tt CHUGUNOV2007} screening routines on the dynamics of the propagating flame in XRBs. Figure \ref{fig:screen_temp} shows the temperature for {\tt aprox13} and {\tt aprox13\_chu} at $t = 50$ ms, which suggests that there are no significant differences in the flame structure between the two models. Figure \ref{fig:screen_gamma_50ms} and \ref{fig:screen_gamma_100ms} show $\Gamma_e$, a measure to determine the approximate screening regime for the flame, at $t = 50$ ms and $t = 100$ ms. 

Although a general trend of decreasing $\Gamma_e$ is observed as the flame progress, it is noteworthy that there are more fusion processes involving heavier nuclei at later stages of burning. Assuming triple-$\alpha$ process, it can inferred that the Coulomb coupling parameter, $\Gamma \sim 0.05$ and $0.025$ for $t = 50$ ms and $t = 100$ ms. Meanwhile, for oxygen burning at $t = 50$ ms and 100 ms, $\Gamma \sim 0.48$ and $0.32$, respectively. These results indicate that helium burning occurs in the weak screening regime, while heavier nuclei burning processes occur in the intermediate screening regime and gradually transition into the weak regime.

\begin{figure*}
\centering
\plotone{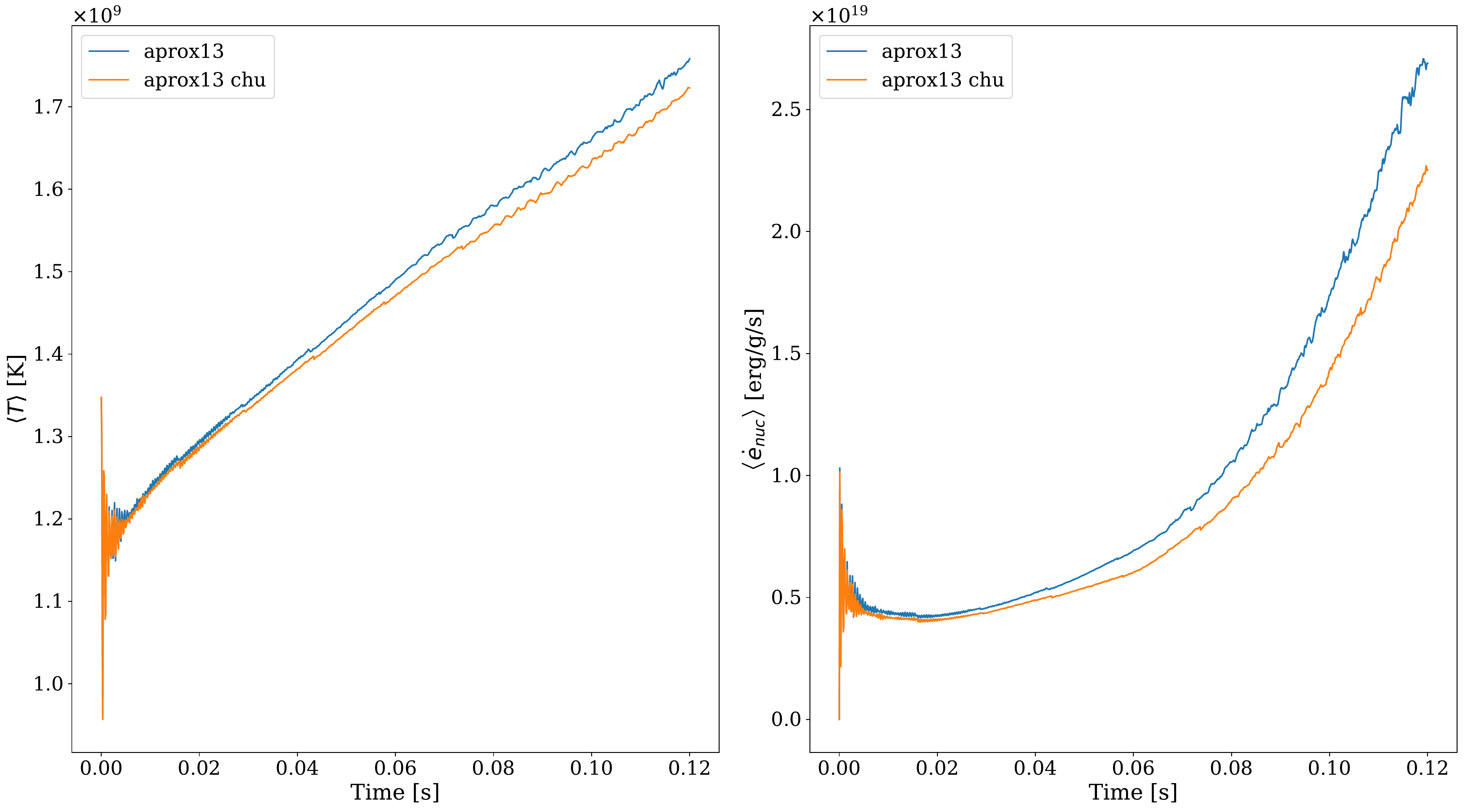}
\caption{\label{fig:screen_profile} Time profiles showing the weighted temperature and energy generation rate of the burning front for {\tt aprox13} and {\tt aprox13\_chu}.}
\end{figure*}

The comparison between {\tt aprox13} and {\tt aprox13\_chu} regarding the weighted temperature and energy generate rate is illustrated in Figure \ref{fig:screen_profile}. This plot illustrates that overall the nuclear energy generation rate of {\tt aprox13} is higher and exhibits a faster increasing rate compared to {\tt aprox13\_chu}. At $t \sim 10$ ms, when the temperature of the two models are approximately equal, the difference in the nuclear energy generation rate indicates that {\tt SCREEN5} provides a stronger screening effect during the initial flame propagation phase compared to {\tt CHUGUNOV2007}.


This finding is in agreement with the results from \cite{Chugunov_2007}, where it was shown that the screening effect calculated by \cite{Alastuey_Jancovici:1978} is always greater than \cite{Chugunov_2007} when $\Gamma \lesssim 40$ (See Figure 4 in \cite{Chugunov_2007}). 
It should be emphasized that {\tt SCREEN5} uses calculation routines from \cite{Alastuey_Jancovici:1978} to formulate the intermediate screening function along with the calculations from \cite{Graboske_1973}. Therefore, reactions that fall within the intermediate screening regime experience a slightly higher screening effect from {\tt SCREEN5} compared to {\tt CHUGUNOV2007}. Consequently, the overall energy generation rates in the early time period are slightly higher for {\tt aprox13} compared to {\tt aprox13\_chu}. This phenomenon accelerates the increasing rate of temperature and the energy generation rate in the later stages of burning, amplifying the discrepancy between these two models as time progresses.


\begin{figure*}
\centering
\plotone{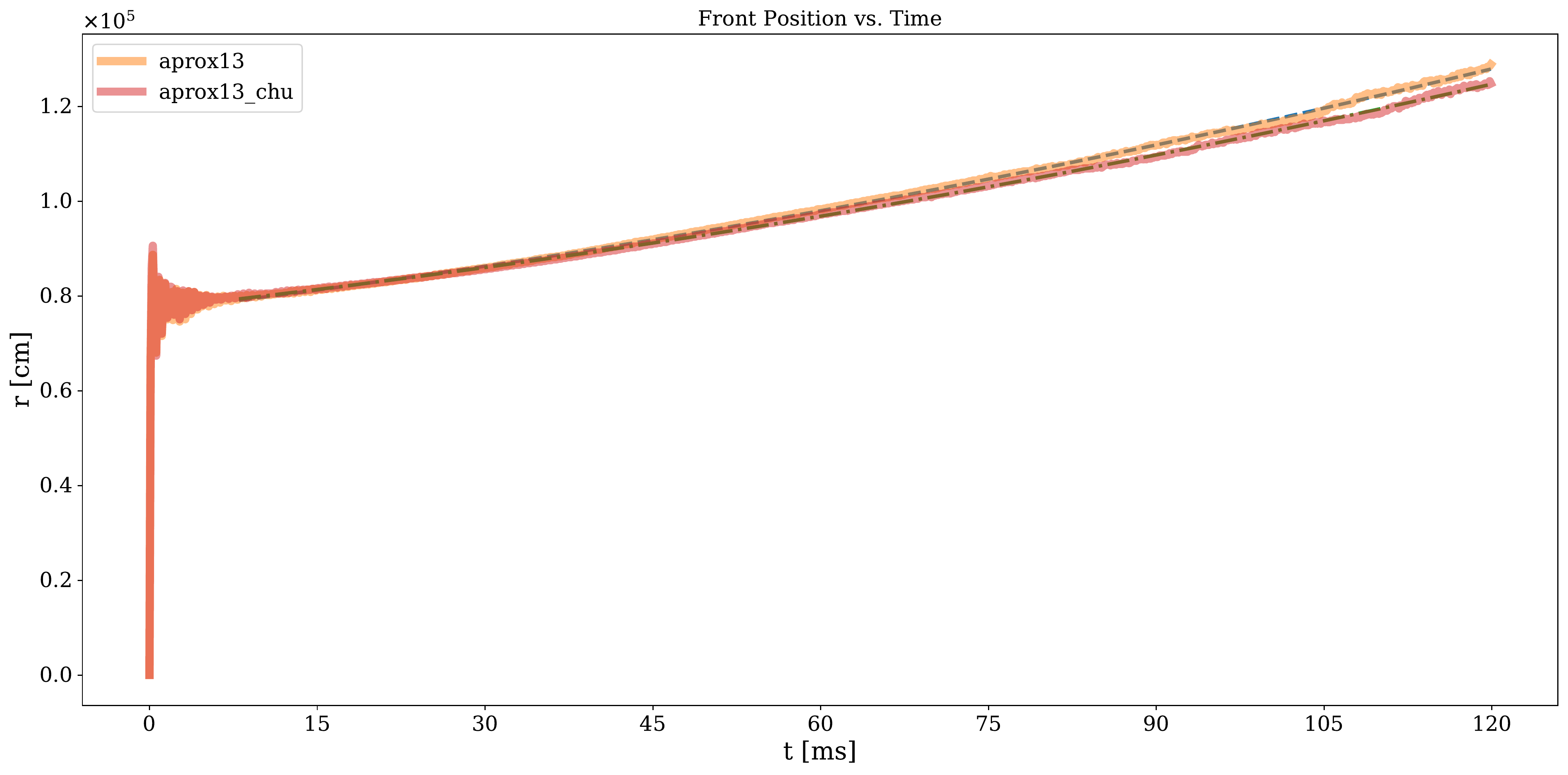}
\caption{\label{fig:screen_front} Flame front position as a function of time for {\tt aprox13} and {\tt aprox13\_chu}. The dashed lines are the fitted curves using Eq. \ref{Eq:quadratic_fit}.}
\end{figure*}

Figure \ref{fig:screen_front} depicts the evolution of the flame front position over time. It is observed that the flame speed for {\tt aprox13\_chu} is slower compared to {\tt aprox13} due to the smaller value of $\dot{e}_{\textrm{nuc}}$ at later times. However, the influence of the two screening methods on the overall flame dynamics is minimal. After fitting the data using Eq. \ref{Eq:quadratic_fit}, the fitted parameters and the instantaneous velocity at various times are shown in Table \ref{Tab:network_fitted_param}.

 \subsection{Time Evolution Method Comparison}\label{Sec:result_integration}



\begin{figure*}
\centering
\plotone{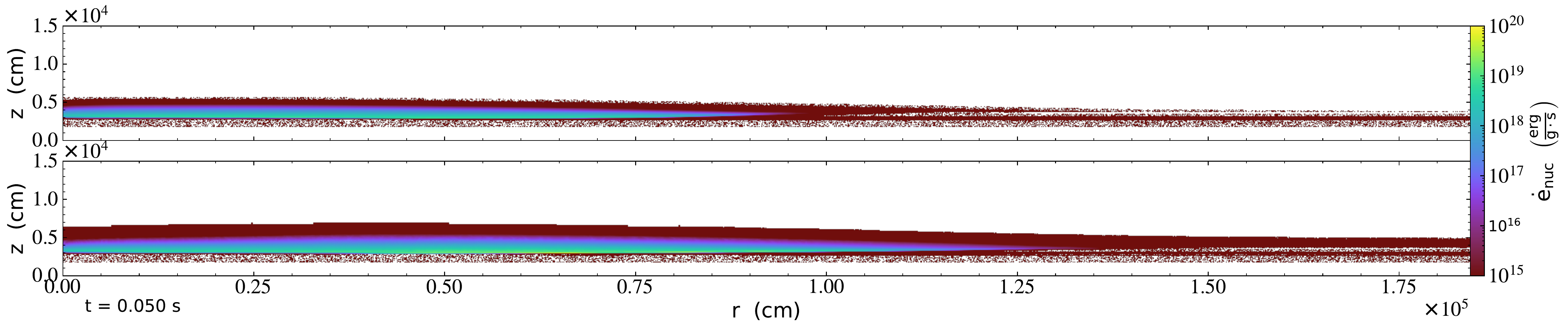}
\caption{\label{fig:integration_enuc} Slice plots comparing $\dot{e}_{\textrm{nuc}}$ for {\tt aprox13\_sdc} (top panel) and {\tt subch\_full\_sdc} (bottom panel) at $t = 50$ ms.}
\end{figure*}

Finally, we compare the simplified-SDC scheme with the traditional Strang-splitting. The overall results using the simplified-SDC scheme are nearly identical to the models that employed Strang-splitting.
One consequence of the strong coupling between advection and reactions is, that to make the energy generation plot, we need to derive the nuclear energy release by subtracting off the advection contribution over a timestep.  In regions where there is not much burning, roundoff error can introduce some noise into the energy generation plot, which is seen as the multiple dotted regions in the $\dot{e}_{\textrm{nuc}}$ plot for {\tt aprox13\_sdc} and {\tt subch\_full\_sdc}.  This is simply an artifact of how we do
the derivation.

\begin{figure*}
\centering
\plotone{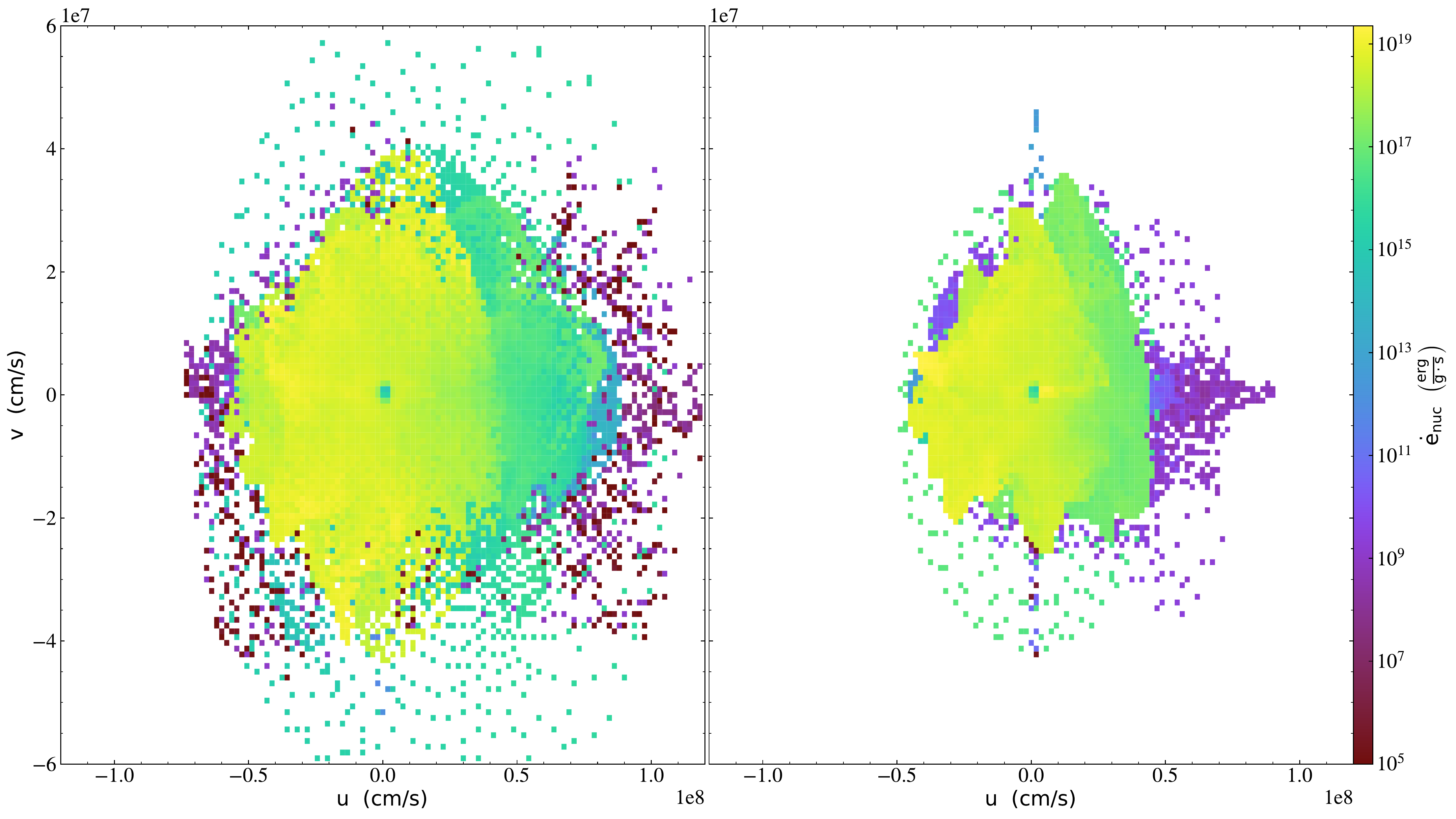}
\caption{\label{fig:integration_u_v_enuc} $u-v$ phase plots for {\tt aprox13\_sdc} (left panel) and {\tt subch\_full\_sdc} (right panel) at $t = 100$ms.}
\end{figure*}

The main advantage of the simplified-SDC integration lies in its stronger coupling between the reaction and hydrodynamics. This feature is demonstrated in the $u$-$v$ phase plot for {\tt aprox13\_sdc} and {\tt subch\_full\_sdc} (Figure \ref{fig:integration_u_v_enuc}). Comparing to the $u$-$v$ phase plots for {\tt aprox13} and {\tt subch\_full} at $100$ ms (Figure \ref{fig:network_u_v_enuc}), Figure \ref{fig:integration_u_v_enuc} shows a smoother blend in the border of the regions with higher and lower $\dot{e}_{\textrm{nuc}}$, corresponding to negative and positive $u$, respectively. The smoother transition suggests this scheme is capable to resolve the flame in a volatile condition.

As shown in \citet{castro_simple_sdc}, in regions where the burning
is vigorous, the simplified-SDC method provides a better solution
than Strang splitting.  However, the XRB flame we simulate here
are not very demanding, so the benefit is minimal.  
As a result, the two simplified-SDC simulations were also more
computationally-expensive than their Strang counterparts.
This differs than the case in \citet{castro_simple_sdc} where 
the simplified-SDC algorithm reduced computational expenses in extreme thermodynamic conditions by mitigating the stiffness of solving reaction equations.

\section{Summary}

We explored the sensitivity of an XRB flame to the details of the
nuclear physics: size of, and approximations in a reaction network, screening methods, and time-integration strategies. The main
differences observed with reaction network are:

\begin{itemize}
    \item The $(\alpha, \mbox{p})(\mbox{p}, \gamma)$ approximation continues to be an accurate approach in simulating thermonuclear flames in XRBs. Up to $t = 120$ ms, the attained temperature during propagation of the thermonuclear flame is $\lesssim 2.5 \times 10^9$ K, and any minor errors associated with the approximation do not significantly affect the overall flame propagation. This conclusion is supported by the similar profiles observed between {\tt subch\_full} and {\tt subch\_simple} networks.
    
    \item The ${}^{12}\mbox{C}(\mbox{p}, \gamma) {}^{13}\mbox{N}(\alpha, \mbox{p}){}^{16}\mbox{O}$ rates are critical in accurately modelling nuclear burning, nucleosynthesis, and flame propagation in XRBs. At $T \gtrsim 10^9$ K, these reactions dominate over the triple-$\alpha$ and the slow $\alpha$ capture processes from ${}^{12}$C to ${}^{16}$O. This allows an instant depletion of ${}^{12}$C, leading to a burst of energy once the temperature reaches $\sim 1.3 \times 10^9$ K. This finding is consistent with the work of \cite{Weinberg_2006}, which claims a similar effect at $1.2 \times 10^9$ K. Upon incorporating these rates into the network, we have successfully simulated an accelerating phase for the laterally propagating flame.

    \item Even though there is not a significant difference between {\tt aprox13} and {\tt subch\_full\_mod}, {\tt subch\_full\_mod} demonstrates an increasingly higher $\dot{e}_{\textrm{nuc}}$. Given that the $(\alpha, \mbox{p})(\mbox{p}, \gamma)$ approximation is accurate, the additional rates must have gradually increased the overall $\dot{e}_{\textrm{nuc}}$ in the long run.
    Another possibility that caused the disparity between {\tt subch\_full\_mod} and {\tt aprox13} is the utilization of updated rates from {\sf REACLIB} library in {\tt subch\_full\_mod}, whereas {\tt aprox13} didn't employ the most up-to-date rates. It is plausible that the contemporary adjustments made to the various reaction rates have contributed to the discrepancy between the two networks. However, further investigations are necessary to confirm this hypothesis.

    \item  Among the four reaction networks used to simulate He flame propagation in XRBs, the {\tt subch\_simple} network proved to be the most effective. It is the smallest network that captures the initial acceleration of the propagating flame, which drastically alters the overall flame dynamics.
    
\end{itemize}

Comparing the two screening routines, {\tt SCREEN5} and {\tt CHUNGUNOV2007}, we find that {\tt CHUGUNOV2007} has a slightly weaker screening effect in the weak and intermediate screening regimes. As a consequence, the weaker screening effect from {\tt CHUGUNOV2007} leads to a slightly slower flame compared to the {\tt SCREEN5} model. This result matches our expectations and is in agreement with \cite{Chugunov_2007}.

Finally, we investigated the performance of the simplified-SDC scheme in comparison to the traditional Strang-splitting.  For this problem, since the burning is not very vigorous, there is no strong benefit of using simplified-SDC over Strang-splitting.

Overall, this study gives us confidence that, by using the {\tt subch\_simple} network for our future simulations, we can accurately capture the dynamics of the flame. Our next step is to adapt the current simulation methodology to model a full star flame propagation model. A full star flame propagation simulation allows us to explore how flame dynamics changes subject to the geometric influence, such as the variations in Coriolis force. As the flame encounters the strongest Coriolis force at the pole and the weakest at the equator, its behavior can alter significantly depending on its position. Additionally, a full star simulation provides a more precise estimate of the time required for the flame to engulf the neutron star, which serves as a better approximation of the XRB's rise time.


\begin{acknowledgements}
\castro\ is open-source and freely available at
\url{http://github.com/AMReX-Astro/Castro}.  The work at Stony Brook was supported by DOE/Office
of Nuclear Physics grant DE-FG02-87ER40317. This research was supported by the Exascale Computing 
Project (17-SC-20-SC), a collaborative effort of the U.S. Department of Energy
Office of Science and the National Nuclear Security Administration. This research used
resources of the National Energy Research Scientific Computing Center,
a DOE Office of Science User Facility supported by the Office of
Science of the U.~S.\ Department of Energy under Contract
No.\ DE-AC02-05CH11231.  This research used resources of the Oak Ridge
Leadership Computing Facility at the Oak Ridge National Laboratory,
which is supported by the Office of Science of the U.S. Department of
Energy under Contract No. DE-AC05-00OR22725, awarded through the DOE
INCITE program.  We thank NVIDIA Corporation for the donation of a
Titan X and Titan V GPU through their academic grant program.  This
research has made use of NASA's Astrophysics Data System Bibliographic
Services.
\end{acknowledgements}

\facilities{NERSC, OLCF}

\software{AMReX \citep{amrex_joss},
          Castro \citep{castro},
          GCC (\url{https://gcc.gnu.org/}),
          linux (\url{https://www.kernel.org/}),
          matplotlib (\citealt{Hunter:2007}, \url{http://matplotlib.org/}),
          NumPy \citep{numpy,numpy2},
          python (\url{https://www.python.org/}),
          valgrind \citep{valgrind},
          VODE \citep{vode},
          yt \citep{yt}
          pynucastro \citep{pynucastro, pynucastro2,the_pynucastro_development_2022_7239007}}


\bibliographystyle{aasjournal}
\bibliography{ws}

\begin{thebibliography}{}
\expandafter\ifx\csname natexlab\endcsname\relax\def\natexlab#1{#1}\fi
\providecommand{\url}[1]{\href{#1}{#1}}
\providecommand{\dodoi}[1]{doi:~\href{http://doi.org/#1}{\nolinkurl{#1}}}
\providecommand{\doeprint}[1]{\href{http://ascl.net/#1}{\nolinkurl{http://ascl.net/#1}}}
\providecommand{\doarXiv}[1]{\href{https://arxiv.org/abs/#1}{\nolinkurl{https://arxiv.org/abs/#1}}}

\bibitem[{Alastuey \& Jancovici(1978{\natexlab{a}})}]{alastuey:1978}
Alastuey, A., \& Jancovici, B. 1978{\natexlab{a}}, ApJ, 226, 1034,
  \dodoi{10.1086/156681}

\bibitem[{Alastuey \& Jancovici(1978{\natexlab{b}})}]{Alastuey_Jancovici:1978}
---. 1978{\natexlab{b}}, ApJ, 226, 1034, \dodoi{10.1086/156681}

\bibitem[{Almgren {et~al.}(2020)Almgren, Sazo, Bell, Harpole, Katz, Sexton,
  Willcox, Zhang, \& Zingale}]{castro_joss}
Almgren, A., Sazo, M.~B., Bell, J., {et~al.} 2020, Journal of Open Source
  Software, 5, 2513, \dodoi{10.21105/joss.02513}

\bibitem[{Almgren {et~al.}(2010)Almgren, Beckner, Bell, Day, Howell, Joggerst,
  Lijewski, Nonaka, Singer, \& Zingale}]{castro}
Almgren, A.~S., Beckner, V.~E., Bell, J.~B., {et~al.} 2010, ApJ, 715, 1221,
  \dodoi{10.1088/0004-637x/715/2/1221}

\bibitem[{Altamirano {et~al.}(2010)Altamirano, Watts, Linares, Markwardt,
  Strohmayer, \& Patruno}]{Altamirano_2010}
Altamirano, D., Watts, A., Linares, M., {et~al.} 2010, Monthly Notices of the
  Royal Astronomical Society, 409, 1136,
  \dodoi{10.1111/j.1365-2966.2010.17369.x}

\bibitem[{Bhattacharyya \& Strohmayer(2006)}]{Bhattacharyya_2006}
Bhattacharyya, S., \& Strohmayer, T.~E. 2006, The Astrophysical Journal, 636,
  L121, \dodoi{10.1086/500199}

\bibitem[{Bourlioux {et~al.}(2003)Bourlioux, Layton, \&
  Minion}]{bourlioux2003reaction_sdc}
Bourlioux, A., Layton, A.~T., \& Minion, M.~L. 2003, Journal of Computational
  Physics, 189, 651

\bibitem[{Brown {et~al.}(1989)Brown, Byrne, \& Hindmarsh}]{vode}
Brown, P.~N., Byrne, G.~D., \& Hindmarsh, A.~C. 1989, SIAM J. Sci. and Stat.
  Comput., 10, 1038, \dodoi{10.1137/0910062}

\bibitem[{Calder {et~al.}(2007)Calder, Townsley, Seitenzahl, Peng, Messer,
  Vladimirova, Brown, Truran, \& Lamb}]{Calder:2007}
Calder, A.~C., Townsley, D.~M., Seitenzahl, I.~R., {et~al.} 2007, The
  Astrophysical Journal, 656, 313, \dodoi{10.1086/510709}

\bibitem[{Cavecchi {et~al.}(2016)Cavecchi, Levin, Watts, \&
  Braithwaite}]{Cavecchi_2016}
Cavecchi, Y., Levin, Y., Watts, A.~L., \& Braithwaite, J. 2016, Monthly Notices
  of the Royal Astronomical Society, 459, 1259, \dodoi{10.1093/mnras/stw728}

\bibitem[{Cavecchi {et~al.}(2013)Cavecchi, Watts, Braithwaite, \&
  Levin}]{cavecchi:2013}
Cavecchi, Y., Watts, A.~L., Braithwaite, J., \& Levin, Y. 2013, Mon. Not. R.
  Astron Soc., 434, 3526, \dodoi{10.1093/mnras/stt1273}

\bibitem[{Cavecchi {et~al.}(2015)Cavecchi, Watts, Levin, \&
  Braithwaite}]{Cavecchi_2015}
Cavecchi, Y., Watts, A.~L., Levin, Y., \& Braithwaite, J. 2015, Monthly Notices
  of the Royal Astronomical Society, 448, 445, \dodoi{10.1093/mnras/stu2764}

\bibitem[{Chabrier \& Potekhin(1998)}]{Chabrier:1998}
Chabrier, G., \& Potekhin, A.~Y. 1998, Physical Review E, 58, 4941,
  \dodoi{10.1103/physreve.58.4941}

\bibitem[{Chakraborty \& Bhattacharyya(2014)}]{Chakraborty_2014}
Chakraborty, M., \& Bhattacharyya, S. 2014, The Astrophysical Journal, 792, 4,
  \dodoi{10.1088/0004-637x/792/1/4}

\bibitem[{Chugunov \& DeWitt(2009)}]{Chugunov_2009}
Chugunov, A.~I., \& DeWitt, H.~E. 2009, Physical Review C, 80,
  \dodoi{10.1103/physrevc.80.014611}

\bibitem[{Chugunov {et~al.}(2007)Chugunov, DeWitt, \& Yakovlev}]{Chugunov_2007}
Chugunov, A.~I., DeWitt, H.~E., \& Yakovlev, D.~G. 2007, Physical Review D, 76,
  \dodoi{10.1103/physrevd.76.025028}

\bibitem[{Colella(1990)}]{ppmunsplit}
Colella, P. 1990, J. Comput. Phys., 87, 171,
  \dodoi{10.1016/0021-9991(90)90233-q}

\bibitem[{Colella \& Woodward(1984)}]{ppm}
Colella, P., \& Woodward, P.~R. 1984, J. Comput. Phys., 54, 174,
  \dodoi{10.1016/0021-9991(84)90143-8}

\bibitem[{Cumming \& Bildsten(2001)}]{Cumming_2001}
Cumming, A., \& Bildsten, L. 2001, The Astrophysical Journal, 559, L127,
  \dodoi{10.1086/323937}

\bibitem[{Cyburt {et~al.}(2010)Cyburt, Amthor, Ferguson, Meisel, Smith, Warren,
  Heger, Hoffman, Rauscher, Sakharuk, Schatz, Thielemann, \&
  Wiescher}]{reaclib}
Cyburt, R.~H., Amthor, A.~M., Ferguson, R., {et~al.} 2010, ApJS, 189, 240,
  \dodoi{10.1088/0067-0049/189/1/240}

\bibitem[{{Dewitt} {et~al.}(1973){Dewitt}, {Graboske}, \&
  {Cooper}}]{Dewitt_1973}
{Dewitt}, H.~E., {Graboske}, H.~C., \& {Cooper}, M.~S. 1973, \apj, 181, 439,
  \dodoi{10.1086/152061}

\bibitem[{Dutt {et~al.}(2000)Dutt, Greengard, \& Rokhlin}]{dutt2000sdc}
Dutt, A., Greengard, L., \& Rokhlin, V. 2000, BIT Numerical Mathematics, 40,
  241

\bibitem[{E.~Willcox \& Zingale(2018)}]{pynucastro}
E.~Willcox, D., \& Zingale, M. 2018, JOSS, 3, 588, \dodoi{10.21105/joss.00588}

\bibitem[{{Eiden} {et~al.}(2020){Eiden}, {Zingale}, {Harpole}, {Willcox},
  {Cavecchi}, \& {Katz}}]{eiden:2020}
{Eiden}, K., {Zingale}, M., {Harpole}, A., {et~al.} 2020, \apj, 894, 6,
  \dodoi{10.3847/1538-4357/ab80bc}

\bibitem[{Fisker {et~al.}(2008)Fisker, Schatz, \& Thielemann}]{fisker:2008}
Fisker, J.~L., Schatz, H., \& Thielemann, F. 2008, ASTROPHYS J SUPPL S, 174,
  261, \dodoi{10.1086/521104}

\bibitem[{Galloway \& Keek(2020)}]{Galloway_2020_basics_of_xrb}
Galloway, D.~K., \& Keek, L. 2020, in Timing Neutron Stars: Pulsations,
  Oscillations and Explosions (Springer Berlin Heidelberg), 209--262,
  \dodoi{10.1007/978-3-662-62110-3_5}

\bibitem[{Galloway {et~al.}(2008)Galloway, Muno, Hartman, Psaltis, \&
  Chakrabarty}]{galloway:2008}
Galloway, D.~K., Muno, M.~P., Hartman, J.~M., Psaltis, D., \& Chakrabarty, D.
  2008, Astrophysical Journal Supplement Series, The, 179, 360,
  \dodoi{10.1086/592044}

\bibitem[{Graboske {et~al.}(1973)Graboske, Dewitt, Grossman, \&
  Cooper}]{Graboske_1973}
Graboske, H.~C., Dewitt, H.~E., Grossman, A.~S., \& Cooper, M.~S. 1973, ApJ,
  181, 457, \dodoi{10.1086/152062}

\bibitem[{Gupta {et~al.}(2007)Gupta, Brown, Schatz, Moller, \&
  Kratz}]{Gupta_2007}
Gupta, S., Brown, E.~F., Schatz, H., Moller, P., \& Kratz, K.-L. 2007, The
  Astrophysical Journal, 662, 1188, \dodoi{10.1086/517869}

\bibitem[{{Harpole} {et~al.}(2021){Harpole}, {Ford}, {Eiden}, {Zingale},
  {Willcox}, {Cavecchi}, \& {Katz}}]{harpole:2021}
{Harpole}, A., {Ford}, N.~M., {Eiden}, K., {et~al.} 2021, \apj, 912, 36,
  \dodoi{10.3847/1538-4357/abee87}

\bibitem[{Hunter(2007)}]{Hunter:2007}
Hunter, J.~D. 2007, Comput. Sci. Eng., 9, 90, \dodoi{10.1109/mcse.2007.55}

\bibitem[{{Itoh} {et~al.}(1996){Itoh}, {Hayashi}, {Nishikawa}, \&
  {Kohyama}}]{neutrino}
{Itoh}, N., {Hayashi}, H., {Nishikawa}, A., \& {Kohyama}, Y. 1996, \apjs, 102,
  411, \dodoi{10.1086/192264}

\bibitem[{Itoh {et~al.}(1979)Itoh, Totsuji, Ichimaru, \& Dewitt}]{itoh:1979}
Itoh, N., Totsuji, H., Ichimaru, S., \& Dewitt, H.~E. 1979, ApJ, 234, 1079,
  \dodoi{10.1086/157590}

\bibitem[{Jancovici(1977)}]{jancovici:1977}
Jancovici, B. 1977, Journal of Statistical Physics, 17, 357

\bibitem[{Johnston {et~al.}(2018)Johnston, Heger, \& Galloway}]{Johnston_2018}
Johnston, Z., Heger, A., \& Galloway, D.~K. 2018, Monthly Notices of the Royal
  Astronomical Society, 477, 2112, \dodoi{10.1093/mnras/sty757}

\bibitem[{Johnston {et~al.}(2020)Johnston, Heger, \& Galloway}]{Johnston_2020}
---. 2020, Monthly Notices of the Royal Astronomical Society, 494, 4576,
  \dodoi{10.1093/mnras/staa1054}

\bibitem[{Kaaret {et~al.}(2007)Kaaret, Prieskorn, in~’t Zand, Brandt, Lund,
  Mereghetti, Götz, Kuulkers, \& Tomsick}]{Kaaret_2007}
Kaaret, P., Prieskorn, Z., in~’t Zand, J. J.~M., {et~al.} 2007, The
  Astrophysical Journal, 657, L97, \dodoi{10.1086/513270}

\bibitem[{Karakas {et~al.}(2008)Karakas, Lee, Lugaro, Görres, \&
  Wiescher}]{Karakas_2008}
Karakas, A.~I., Lee, H.~Y., Lugaro, M., Görres, J., \& Wiescher, M. 2008, The
  Astrophysical Journal, 676, 1254, \dodoi{10.1086/528840}

\bibitem[{Koike {et~al.}(2004)Koike, aki Hashimoto, Kuromizu, \& ichirou
  Fujimoto}]{Koike_2004}
Koike, O., aki Hashimoto, M., Kuromizu, R., \& ichirou Fujimoto, S. 2004, The
  Astrophysical Journal, 603, 242, \dodoi{10.1086/381354}

\bibitem[{{Kuulkers, E.}(2002)}]{Kuulkers_2002}
{Kuulkers, E.} 2002, A\&A, 383, L5, \dodoi{10.1051/0004-6361:20011811}

\bibitem[{Meisel(2018)}]{Meisel_2018}
Meisel, Z. 2018, The Astrophysical Journal, 860, 147,
  \dodoi{10.3847/1538-4357/aac3d3}

\bibitem[{Miller \& Colella(2002)}]{millercolella:2002}
Miller, G., \& Colella, P. 2002, J. Comput. Phys., 183, 26,
  \dodoi{10.1006/jcph.2002.7158}

\bibitem[{Nethercote \& Seward(2007)}]{valgrind}
Nethercote, N., \& Seward, J. 2007, in Proceedings of the 2007 ACM SIGPLAN
  conference on Programming language design and implementation - PLDI '07, PLDI
  '07 (New York, NY, USA: ACM Press), 89{\textendash}100,
  \dodoi{10.1145/1250734.1250746}

\bibitem[{Newton {et~al.}(2007)Newton, Iliadis, Champagne, Coc, Parpottas, \&
  Ugalde}]{Newton2007}
Newton, J.~R., Iliadis, C., Champagne, A.~E., {et~al.} 2007, Phys. Rev. C, 75,
  045801, \dodoi{10.1103/PhysRevC.75.045801}

\bibitem[{Oliphant(2007)}]{numpy}
Oliphant, T.~E. 2007, Comput. Sci. Eng., 9, 10, \dodoi{10.1109/mcse.2007.58}

\bibitem[{Parikh {et~al.}(2013)Parikh, Jos{\'{e}}, Sala, \&
  Iliadis}]{Parikh_2013}
Parikh, A., Jos{\'{e}}, J., Sala, G., \& Iliadis, C. 2013, Progress in Particle
  and Nuclear Physics, 69, 225, \dodoi{10.1016/j.ppnp.2012.11.002}

\bibitem[{{Shara}(1982)}]{Shara_1982}
{Shara}, M.~M. 1982, \apj, 261, 649, \dodoi{10.1086/160376}

\bibitem[{Shen \& Bildsten(2009)}]{Shen_2009}
Shen, K.~J., \& Bildsten, L. 2009, The Astrophysical Journal, 699, 1365,
  \dodoi{10.1088/0004-637x/699/2/1365}

\bibitem[{Smith {et~al.}(2023)Smith, Johnson, Chen, Eiden, Willcox, Boyd, Cao,
  DeGrendele, \& Zingale}]{pynucastro2}
Smith, A.~I., Johnson, E.~T., Chen, Z., {et~al.} 2023, The Astrophysical
  Journal, 947, 65, \dodoi{10.3847/1538-4357/acbaff}

\bibitem[{Smith {et~al.}(1997)Smith, Morgan, \& Bradt}]{Smith_1997}
Smith, D.~A., Morgan, E.~H., \& Bradt, H. 1997, The Astrophysical Journal, 479,
  L137, \dodoi{10.1086/310604}

\bibitem[{Spitkovsky {et~al.}(2002)Spitkovsky, Levin, \&
  Ushomirsky}]{Spitkovsky_2002}
Spitkovsky, A., Levin, Y., \& Ushomirsky, G. 2002, The Astrophysical Journal,
  566, 1018, \dodoi{10.1086/338040}

\bibitem[{Strang(1968)}]{strang:1968}
Strang, G. 1968, SIAM J. Numer. Anal., 5, 506, \dodoi{10.1137/0705041}

\bibitem[{Strohmayer {et~al.}(2009)Strohmayer, Zhang, Swank, Smale, Titarchuk,
  Day, \& Lee}]{strohmayer_2009}
Strohmayer, E., Zhang, W., Swank, H., {et~al.} 2009, The Astrophysical Journal
  Letters, 469, L9, \dodoi{10.1086/310261}

\bibitem[{the~pynucastro development {et~al.}(2022)the~pynucastro development,
  Boyd, Cao, Chen, Eiden, Johnson, Li, Smith~Clark, Willcox, \&
  Zingale}]{the_pynucastro_development_2022_7239007}
the~pynucastro development, Boyd, B., Cao, L., {et~al.} 2022,
  pynucastro/pynucastro: pynucastro 2.0.2, 2.0.2,  Zenodo,
  \dodoi{10.5281/zenodo.7239007}

\bibitem[{the StarKiller Microphysics Development~Team {et~al.}(2019)the
  StarKiller Microphysics Development~Team, Bishop, Fields, Jacobs, Katz,
  Malone, Timmes, Willcox, \& Zingale}]{starkiller}
the StarKiller Microphysics Development~Team, Bishop, A., Fields, C.~E.,
  {et~al.} 2019, starkiller-astro/Microphysics: 19.04,
  \dodoi{10.5281/zenodo.2620545}

\bibitem[{Timmes(2000)}]{Timmes00}
Timmes, F.~X. 2000, ApJ, 528, 913, \dodoi{10.1086/308203}

\bibitem[{{Timmes}(2019)}]{timmes_aprox13}
{Timmes}, F.~X. 2019

\bibitem[{Timmes \& Swesty(2000)}]{timmes_swesty:2000}
Timmes, F.~X., \& Swesty, F.~D. 2000, ASTROPHYS J SUPPL S, 126, 501,
  \dodoi{10.1086/313304}

\bibitem[{Turk {et~al.}(2010)Turk, Smith, Oishi, Skory, Skillman, Abel, \&
  Norman}]{yt}
Turk, M.~J., Smith, B.~D., Oishi, J.~S., {et~al.} 2010, ApJS, 192, 9,
  \dodoi{10.1088/0067-0049/192/1/9}

\bibitem[{van~der Walt {et~al.}(2011)van~der Walt, Colbert, \&
  Varoquaux}]{numpy2}
van~der Walt, S., Colbert, S.~C., \& Varoquaux, G. 2011, Comput. Sci. Eng., 13,
  22, \dodoi{10.1109/mcse.2011.37}

\bibitem[{{Wallace} {et~al.}(1982){Wallace}, {Woosley}, \&
  {Weaver}}]{Wallace:1982}
{Wallace}, R.~K., {Woosley}, S.~E., \& {Weaver}, T.~A. 1982, \apj, 258, 696,
  \dodoi{10.1086/160119}

\bibitem[{Weinberg \& Bildsten(2007)}]{Weinberg2007}
Weinberg, N.~N., \& Bildsten, L. 2007, ApJ, 670, 1291, \dodoi{10.1086/522111}

\bibitem[{{Weinberg} {et~al.}(2006){Weinberg}, {Bildsten}, \&
  {Schatz}}]{Weinberg_2006}
{Weinberg}, N.~N., {Bildsten}, L., \& {Schatz}, H. 2006, \apj, 639, 1018,
  \dodoi{10.1086/499426}

\bibitem[{Woosley {et~al.}(2004{\natexlab{a}})Woosley, Wunsch, \&
  Kuhlen}]{woosley-ignition}
Woosley, S.~E., Wunsch, S., \& Kuhlen, M. 2004{\natexlab{a}}, ApJ, 607, 921,
  \dodoi{10.1086/383530}

\bibitem[{Woosley {et~al.}(2004{\natexlab{b}})Woosley, Heger, Cumming, Hoffman,
  Pruet, Rauscher, Fisker, Schatz, Brown, \& Wiescher}]{Woosley_2004}
Woosley, S.~E., Heger, A., Cumming, A., {et~al.} 2004{\natexlab{b}}, The
  Astrophysical Journal Supplement Series, 151, 75, \dodoi{10.1086/381533}

\bibitem[{Yakovlev {et~al.}(2006)Yakovlev, Gasques, Afanasjev, Beard, \&
  Wiescher}]{Yakovlev_2006}
Yakovlev, D.~G., Gasques, L.~R., Afanasjev, A.~V., Beard, M., \& Wiescher, M.
  2006, Physical Review C, 74, \dodoi{10.1103/physrevc.74.035803}

\bibitem[{Zhang {et~al.}(2019)Zhang, Almgren, Beckner, Bell, Blaschke, Chan,
  Day, Friesen, Gott, Graves, Katz, Myers, Nguyen, Nonaka, Rosso, Williams, \&
  Zingale}]{amrex_joss}
Zhang, W., Almgren, A., Beckner, V., {et~al.} 2019, JOSS, 4, 1370,
  \dodoi{10.21105/joss.01370}

\bibitem[{{Zingale} {et~al.}(2023){Zingale}, {Eiden}, \& {Katz}}]{xrb_flame_3d}
{Zingale}, M., {Eiden}, K., \& {Katz}, M. 2023, arXiv e-prints,
  arXiv:2303.17077, \dodoi{10.48550/arXiv.2303.17077}

\bibitem[{Zingale {et~al.}(2019)Zingale, Katz, Bell, Minion, Nonaka, \&
  Zhang}]{castro-sdc}
Zingale, M., Katz, M.~P., Bell, J.~B., {et~al.} 2019, ApJ, 886, 105,
  \dodoi{10.3847/1538-4357/ab4e1d}

\bibitem[{Zingale {et~al.}(2022)Zingale, Katz, Nonaka, \&
  Rasmussen}]{castro_simple_sdc}
Zingale, M., Katz, M.~P., Nonaka, A., \& Rasmussen, M. 2022, The Astrophysical
  Journal, 936, 6, \dodoi{10.3847/1538-4357/ac8478}

\bibitem[{Zingale {et~al.}(2021)Zingale, Katz, Willcox, \&
  Harpole}]{castro_strang}
Zingale, M., Katz, M.~P., Willcox, D.~E., \& Harpole, A. 2021, Research Notes
  of the AAS, 5, 71, \dodoi{10.3847/2515-5172/abf3cb}

\end{thebibliography}

\end{document}